\documentclass[10pt,twocolumn]{IEEEtran}

\usepackage{enumitem}
\usepackage{amsmath,graphics,amssymb,epsfig,subfigure,color,cite}

\usepackage{amsthm}
\theoremstyle{plain}
\newtheoremstyle{mystyle}
  {0mm}
  {0mm}
  {}
  {4mm}
  {\bfseries}
  {:}
  { }
  {\thmname{#1}\thmnumber{ #2}\thmnote{ (#3)}}
\theoremstyle{mystyle}

\usepackage{array}
\usepackage{multirow}
\usepackage{enumerate}
\usepackage{algorithmic}
\usepackage{algorithm}
\usepackage{bm}
\usepackage{float}
\usepackage{makecell}
\usepackage[right=0.625in, left=0.625in, top=0.7in, bottom=1.1in]{geometry}

\renewcommand{\thefigure}{\arabic{figure}}

\floatname{algorithm}{Procedure}

\makeatletter
\newcommand{\vast}{\bBigg@{4.5}}
\newcommand{\Vast}{\bBigg@{7.5}}
\makeatother

\begin{document}

\title{Deep Learning-Assisted Parallel Interference Cancellation for Grant-Free NOMA in Machine-Type Communication}
\author{Yongjeong Oh, Jaehong Jo, Byonghyo Shim, and Yo-Seb Jeon
	    \thanks{Yongjeong Oh, Jaehong Jo, and Yo-Seb Jeon are with the Department of Electrical Engineering, POSTECH, Pohang, Gyeongbuk 37673, Republic of Korea (e-mails: \{yongjeongoh, jaehongjo, yoseb.jeon\}@postech.ac.kr).}
        \thanks{Byonghyo Shim is with the Institute of New Media and Communications
and the Department of Electrical and Computer Engineering, Seoul National
University, Seoul 08826, Republic of Korea (e-mail: bshim@snu.ac.kr).}
	}
	\vspace{-2mm}	
	
	\maketitle
	\vspace{-12mm}

\begin{abstract} 
In this paper, we present a novel approach for joint activity detection (AD), channel estimation (CE), and data detection (DD) in uplink grant-free non-orthogonal multiple access (NOMA) systems. Our approach employs an iterative and parallel interference removal strategy inspired by parallel interference cancellation (PIC), enhanced with deep learning to jointly tackle the AD, CE, and DD problems. Based on this approach, we develop three PIC frameworks, each of which is designed for either coherent or non-coherence schemes. The first framework performs joint AD and CE using received pilot signals in the coherent scheme. Building upon this framework, the second framework utilizes both the received pilot and data signals for CE, further enhancing the performances of AD, CE, and DD in the coherent scheme. The third framework is designed to accommodate the non-coherent scheme involving a small number of data bits, which simultaneously performs AD and DD. Through joint loss functions and interference cancellation modules, our approach supports end-to-end training, contributing to enhanced performances of AD, CE, and DD for both coherent and non-coherent schemes. Simulation results demonstrate the superiority of our approach over traditional techniques, exhibiting enhanced performances of AD, CE, and DD while maintaining lower computational complexity.

\end{abstract}

\begin{IEEEkeywords}
    Grant-free non-orthogonal multiple access, parallel interference cancellation, deep learning, end-to-end training, non-coherent scheme.
\end{IEEEkeywords}
	

\section{Introduction}\label{Sec:Intro}
With the rapid growth of the Internet of Things (IoT), massive machine-type communication (mMTC) has been recognized as one of the core services for 5G and 6G \cite{survey5G,survey6G,survey1,9205230}. While there are a vast number of IoT devices in the network, only a fraction of them are activated at a given time, and they transmit small amounts of sensing, control, and command information. Traditional grant-based orthogonal multiple access will be clearly inefficient because it requires a complicated handshaking mechanism, even when a device transmits a small number of data bits \cite{WJS,survey3}.

To overcome the problem, grant-free non-orthogonal multiple access (NOMA) has emerged as a promising solution for mMTC \cite{8533378,9364871,survey3,SBH,survey2,CS1,Model_driven,9634107}.
Overall, grant-free NOMA systems can be classified into two classes: (i) the coherent scheme and (ii) the non-coherent scheme. In the coherent scheme, each active device transmits a unique spreading sequence as a pilot signal, followed by data signals, without engaging in a resource-intensive handshaking process \cite{survey2}. Upon receiving the superimposed signal from the active devices, the base station (BS) identifies their activities and channels through activity detection (AD) and channel estimation (CE), leveraging the knowledge of their spreading sequences. After AD and CE, the BS then performs data detection (DD) based on the estimated activities and channels. 

In the non-coherent scheme, each device is assigned a set of spreading sequences and, upon activation, transmits one of these sequences based on the data bits \cite{AMP_NC1,CS1,Model_driven}. After receiving the superimposed signal, the BS performs the joint estimation of device activities and data bits by identifying the transmitted spreading sequences. Often, the choice of coherent and non-coherent types depends on the number of data bits \cite{CS1}. In particular, in mMTC scenarios where the number of data bits is small, the non-coherent scheme is preferred. Whereas, when the number of data bits is large, the coherent scheme is popularly used since the non-coherent scheme suffers from expensive computational complexity. 

The aforementioned grant-free NOMA schemes do not require an explicit handshaking process, which leads to a significant reduction in the communication latency. Furthermore, by sharing the same time-frequency resources among multiple devices, better utilization of communication resources can be achieved. 
Despite these advantages, they also face several problems.
For example, since the IoT devices outnumber the communication resource units, it is very difficult to assign orthogonal spreading sequences to each device, which causes collisions among devices and subsequent performance degradation \cite{survey2,9189944}. 
\subsection{Prior Works}
In recent years, to address the challenge we mentioned, a compressed sensing (CS) approach has been popularly used for AD and CE in the coherent scheme \cite{BCS,AMPNN,AMP_sim,Sim_set} and AD and DD in the non-coherent scheme \cite{AMP_NC1,CS1,Model_driven}. 
This approach exploits the sparse activity patterns exhibited by numerous devices. It also provides performance guarantees for AD, CE, and DD in an asymptotic regime where the number of devices and the length of spreading sequences approach infinity while maintaining a fixed ratio between them \cite{CS1}. Despite these advantages, CS-based techniques have certain limitations and practical issues to address. 
First, since the length of the spreading sequence is finite due to the limited coherence time, the performance guarantees obtained in the asymptotic regime cannot be applied. 
This challenge becomes particularly noticeable in critical or low-latency mMTC scenarios, where a short spreading sequence is essential to satisfy the latency requirements \cite{CmMTC1, JCH_PJH}. 
Second, the restricted isometric property (RIP), which is a fundamental requirement for ensuring the accuracy of CS recovery techniques, might be violated when the number of active devices is large \cite{Cluster}. 
These inherent limitations of the CS approach can result in significant performance degradation for AD, CE, and DD, thereby constraining its broader applicability to diverse grant-free NOMA scenarios.  

Successive interference cancellation (SIC) has been recognized as a promising alternative to tackle AD, CE, and DD problems in grant-free NOMA \cite{SIC1,SIC2,SIC3}. The main advantage of the SIC lies in its ability to iteratively eliminate interference, thereby enhancing the overall system performance. However, in the absence of accurate knowledge about activity and channel gain, determining the proper cancellation order becomes challenging. 
This difficulty may result in significant error propagation, leading to a degradation in overall performance \cite{SIC3}. 
Parallel interference cancellation (PIC) can address the aforementioned problem as this approach does not require a prior knowledge about the channel gain order \cite{PIC_CE,DeepNOMA,IC_survey}. Furthermore, the PIC approach can achieve lower latency compared to the SIC approach \cite{IC_survey}. 
Despite these benefits, most PIC techniques to date do not fully account for certain unique characteristics of grant-free NOMA systems, such as non-orthogonal spreading sequences and sparse activity patterns.
Neglecting these characteristics can potentially lead to error propagation, thereby limiting the performance gains offered by PIC.  
Moreover, to the best of our knowledge, none of the existing works have investigated the advantages of a PIC approach for AD, CE, and DD in grant-free NOMA, despite its potential advantages in addressing the limitations of the CS and SIC approaches.

\subsection{Contributions}
In this paper, we propose a deep learning (DL)-assisted PIC approach for grant-free NOMA systems. Specifically, we develop three PIC frameworks based on the proposed approach: (i) the pilot-only PIC, (ii) the data-aided PIC, and (iii) the non-coherent PIC.
The key idea of these frameworks is to eliminate inter-user interference in an iterative and parallel manner inspired by PIC, while jointly optimizing the AD, CE, and DD performances via end-to-end training facilitated by DL.
Notably, our frameworks surpass existing techniques in both performance and computational efficiency without requiring prior knowledge of channel gain orders and device activities.
The major contributions of this paper can be summarized as follows:
\begin{itemize}
    \item We present a DL-assisted pilot-only PIC framework for the coherent scheme in grant-free NOMA. 
    The proposed framework comprises multiple stages, each incorporating three modules: CE, IC, and AD modules. Each module performs its specified task, while simultaneously collaborating to enhance the overall performance of the grant-free NOMA system. Furthermore, the proposed framework supports end-to-end training, which can improve overall performance by training complex relationships and dependencies between different modules. 


    \item We extend the pilot-only PIC framework into the data-aided PIC framework. The most prominent feature of this framework is that it utilizes the received data signals as well as the received pilot signals for CE. To this end, in the data-aided framework, both CE and DD modules are included as a part of end-to-end learning, while PIC is applied to both the received pilot and data signals.
    By incorporating additional information for the CE module, this framework not only improves CE performance but also enhances AD and DD performances compared to the pilot-only PIC framework. 

    \item We extend the pilot-only PIC framework into the non-coherent PIC framework. 
    The distinctive feature of this framework is its ability to simultaneously perform AD and DD, using a single unified module. By doing so, the non-coherent PIC framework effectively addresses the inherent inefficiency of coherent schemes, which necessitate the transmission of pilot signals even when dealing with small data bits. 

    
    
    

    
    
    
    \item
    Through extensive numerical evaluation, we demonstrate the superior performance of our frameworks compared to traditional techniques. 
    Our simulation results show that, in a scenario with $20$ devices, $18$-length coherence interval, and $2$ data bits, the proposed pilot-only PIC framework achieves more than a $1.25$-fold decrease in both AD and DD errors and a $3.3$dB decrease in CE error compared to the AMP technique.
    Furthermore, it is demonstrated that, in the same scenario above, the proposed data-aided PIC framework achieves more than a twofold decrease in both AD and DD errors and a $6.9$dB decrease in CE error compared to the AMP technique. Moreover, the proposed pilot-only and data-aided PIC frameworks achieves these improvements with less computational complexities. The effectiveness of the proposed non-coherent PIC framework is also verified when the number of data bits is small. The computational complexities of the proposed frameworks are also analyzed in terms of simulation time. 


\end{itemize}

\subsubsection*{Organization}
The remainder of the paper is organized as follows. In
Section~\ref{Sec:System}, we first introduce an uplink grant-free NOMA system with coherent and non-coherent schemes. We then summarize the key challenges inherent in this system. In Section~\ref{Sec:Comp}, we present the proposed pilot-only PIC framework that tackles the key challenges in grant-free NOMA. In Section~\ref{Sec:Extend}, we discuss the extensions of the pilot-only PIC framework, focusing on the data-aided PIC and non-coherent PIC frameworks.
In Section~\ref{Sec:Simul}, we provide simulation results that demonstrate the superiority of the proposed frameworks. Finally, in Section~\ref{Sec:Conclusion}, we present our conclusions and future research directions. 

\subsubsection*{Notation}
Upper-case and lower-case boldface letters denote matrices and column vectors, respectively. $\mathbb{E}[\cdot]$ is the statistical expectation and $\mathbb{P}(\cdot)$ is the probability. 
${\sf Re}\{\cdot\}$ and ${\sf Im}\{\cdot\}$ denote real and imaginary components, respectively.
	$\|{\boldsymbol a}\|_2$ and $\|{\boldsymbol a}\|_1$ are the Euclidean norm and Manhattan norm of a real vector ${\boldsymbol a}$, respectively. $\lfloor \cdot \rfloor$ is the floor function. 
     ${\bm 0}_n$ is the vector with all entries equal to zero.
	${\boldsymbol I}_N$ is an $N$ by $N$ identity matrix.

\section{System Model}\label{Sec:System}
In this section, we first describe the uplink grant-free NOMA system with two communication schemes, and then discuss key challenges behind the practical deployment of these systems.

\begin{figure}[t]
    \centering 
    {\epsfig{file=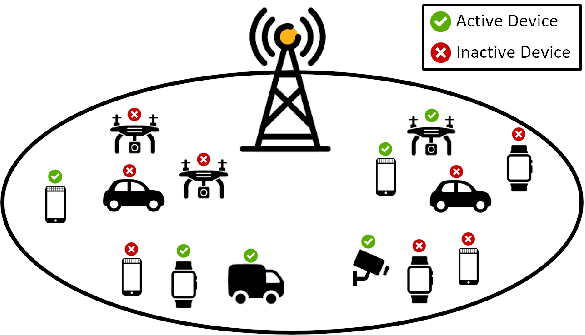,width=7.5cm}}
    \caption{Illustration of the uplink grant-free NOMA system considered in our work.}  \vspace{-2mm}
    \label{fig:Sys}
\end{figure}

\subsection{Uplink Grant-Free NOMA System}\label{Sec:SystemA}
\begin{figure*}[t]
    \centering 
    {\epsfig{file=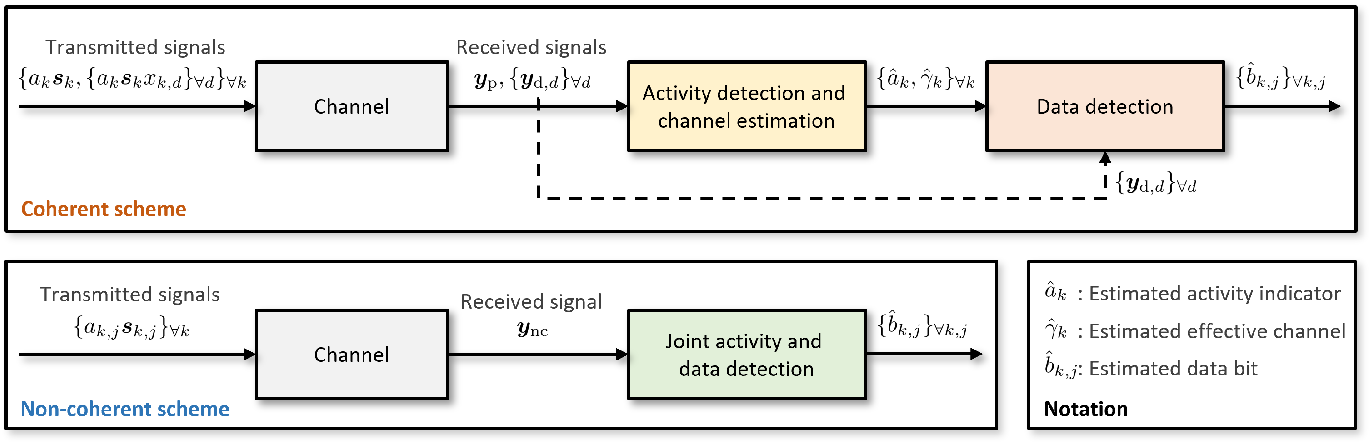,width=15cm}}
    \caption{Overall diagram of two communication schemes in grant-free NOMA: (i) coherent scheme and (ii) non-coherent scheme.}  
    \label{fig:Scheme}
\end{figure*}

We consider an uplink grant-free NOMA system comprising $K$ devices and one BS, as illustrated in Fig.~\ref{fig:Sys}. Without loss of generality, we assume that both the devices and the BS are equipped with a single antenna. 
The channel between device $k$ and the BS is modeled as follows \cite{CS1,Model_driven}:
\begin{align}\label{eq:channel}
    g_k = \sqrt{\beta_k}h_k,
\end{align}
where $\beta_k$ is a large-scale fading coefficient and $h_k\sim\mathcal{CN}(0,1)$ is a small-scale fading channel. The channel $g_k$ remains constant during the transmission of $\tau$ samples. The device $k$ becomes active with a probability of $\epsilon$ and aims to transmit $J$ data bits $[b_1,b_2,\ldots,b_{J}]^{\sf T}\in\{0,1\}^J$ to the BS. 
Under this system, two communication schemes can be considered \cite{survey2, CS1}: (i) coherent scheme and (ii) non-coherent scheme. The overall diagram of these two schemes is illustrated in Fig.~\ref{fig:Scheme}, with detailed explanations provided below.  
\begin{itemize}
    \item {\bf Coherent scheme:} In this scheme, each device $k$ has a unique spreading sequence ${\bm s}_k\in\mathbb{C}^L$, known to the BS. All spreading sequences have unit power, i.e., $\|{\bm s}_k\|_2=1, \forall k$, and their lengths are smaller than the number of devices, i.e., $L<K$. When a certain device $k$ is active, it modulates $J$ data bits to $D$ symbols. The $d$th symbol is represented as $x_{k,d} \in \mathcal{C}$, where $\mathcal{C}$ denotes a constellation set, and $d\in\{1,\ldots,D\}$. Subsequently, active device $k$ spreads $d$th data symbol $x_{k,d} \in \mathcal{C}$ using the spreading sequence, generating ${\bm s}_k x_{k,d}$. During the coherent transmission, each active device $k$ first transmits its spreading sequence ${\bm s}_k$ as a pilot signal for AD and CE, and then transmits the $D$ sequences $\{{\bm s}_k x_{k,d}\}_{\forall d}$ as data signals to the BS. 
    Under this transmission strategy, the sequence length $L$ should be less than $\tau/(D+1)$ due to the limited channel coherence time. The received pilot and data signals at the BS are given by
    \begin{align}\label{eq:received_sig_pilot}
        {\bm y}_{\rm p} &=  \sum_{k=1}^K \rho{\bm s}_k a_k g_k + {\bm n}_{\rm p} = \sum_{k=1}^K {\bm s}_k \gamma_k + {\bm n}_{\rm p},
    \end{align}
    and
    \begin{align}\label{eq:received_sig_data}
        {\bm y}_{{\rm d}, d} &=  \sum_{k=1}^K \rho{\bm s}_k a_k g_k x_{k,d} + {\bm n}_{{\rm d}, d} = \sum_{k=1}^K {\bm s}_k \gamma_k x_{k,d} + {\bm n}_{{\rm d}, d},
    \end{align}
    respectively, where $\rho$ is the transmission power of the devices, $a_k \in \{0,1\}$ is the device activity indicator with $\mathbb{P}[a_k = 1] = \epsilon$ and $\mathbb{P}[a_k = 0] = 1-\epsilon$, and ${\bm n}_{\rm p}$ and ${\bm n}_{{\rm d}, d}$ represent the additive white Gaussian noise (AWGN) at the BS. 
    After receiving the superimposed signals in \eqref{eq:received_sig_pilot} and \eqref{eq:received_sig_data}, the BS estimates the device activity indicators and the effective channels, subsequently detecting the data symbols $\{x_{k,d}\}_{\forall k,d}$.

    \item {\bf Non-coherent scheme:} In this scheme, each device $k$ has a set of spreading sequences $\mathcal{S}_k = \{{\bm s}_{k,1}, {\bm s}_{k,2}, \ldots, {\bm s}_{k,2^J}\}$, where each ${\bm s}_{k,j}\in\mathbb{C}^L$ is known to the BS and has unit power, i.e., $\|{\bm s}_{k,j}\|_2 = 1, \forall k,j$. When a device $k$ is active, it chooses one of these sequences based on the $J$ data bits. Then, the active device $k$ transmits this selected spreading sequence to the BS. Under the non-coherent transmission strategy, the sequence length $L$ should be less than $\tau$. The received signal at the BS is given by
    \begin{align}\label{eq:received_sig_nc}
        {\bm y}_{\rm nc} &=  \sum_{k=1}^K \sum_{j=1}^{2^J}\rho{\bm s}_{k,j} a_{k,j} g_k + {\bm n}_{\rm nc} \nonumber\\&= \sum_{k=1}^K \sum_{j=1}^{2^J} {\bm s}_{k,j} \gamma_{k,j} + {\bm n}_{\rm nc},
    \end{align}
    where ${\bm n}_{\rm nc}$ is the AWGN at the BS, $a_{k,j} \in \{0,1\}$ indicates whether or not $j$th sequence of device $k$ is transmitted, and satisfies
    \begin{align}\label{eq:true_akj_NC}
        \sum_{j=1}^{2^J}{a}_{k,j} = 
        \begin{cases} 
            {1,}&{\text{with probability }\epsilon},\\ 
            {0,}&{\text{with probability }1-\epsilon}. 
        \end{cases}
    \end{align}
    In this scheme, we use the notation $a_k = \sum_{j=1}^{2^J}{a}_{k,j} \in \{0,1\}$ to represent the device activity indicator, which aligns with the terminology used in the coherent scheme. After receiving the superimposed signal in \eqref{eq:received_sig_nc}, the BS estimates $\{a_{k,j}\}_{\forall k,j}$, and demodulates them to obtain data bits. 
\end{itemize}

In the non-coherent scheme, the BS should take into account $2^JK$ sequences, which are far longer than the $K$ sequences in the coherent scheme. Consequently, the non-coherent scheme exhibits expensive computational complexity, particularly when the number $J$ of data bits is large. Despite this computational overhead, the non-coherent scheme offers superior performance in scenarios where $J$ is small, as reported in \cite{CS1}. 

Let $\hat{a}_k$, $\hat{\gamma}_k$, $\hat{b}_{k,j}$ be the estimated activity indicator, effective channel, and $j$th data bit of device $k$, respectively. Then, the performance of AD is evaluated using the false alarm probability $P_{\rm fa} = \mathbb{P}[\hat{a}_k = 1|a_k = 0]$ and the miss detection probability $P_{\rm md} = \mathbb{P}[\hat{a}_k = 0|a_k = 1]$ \cite{DL_based_UAD_CE}.
The average error probability is given by
\begin{align}\label{eq:P_ALL}
   P_{\rm err} = (1-\epsilon)P_{\rm fa} + \epsilon P_{\rm md}. 
\end{align}
If $P_{\rm fa}$ and $P_{\rm md}$ are equal, one can trivially have that $P_{\rm err} = P_{\rm fa} = P_{\rm md}$. 
The performance of CE is evaluated by the normalized mean-squared error (NMSE) defined as 
\begin{align}\label{eq:CE}
   {\rm NMSE} = \mathbb{E}\left[\frac{\sum_{k=1}^K|\hat{\gamma}_k - {\gamma}_k|^2}{\sum_{k=1}^K|\gamma_k|^2}\right].
\end{align}
The performance of DD is assessed using the bit error rate (BER) defined as
\begin{align}\label{eq:BER}
   {\rm BER} = 1 - \big(&\mathbb{P}[\hat{b}_{k,j}=0,\hat{a}_k=1|b_{k,j}=0,a_k=1] \nonumber \\ 
   +\,&\mathbb{P}[\hat{b}_{k,j}=1,\hat{a}_k=1|b_{k,j}=1,a_k=1] \nonumber \\
   +\,&\mathbb{P}[\hat{a}_k=0|a_k=0]\big), 
\end{align}
where, if $a_k=0$ and $\hat{a}_k=1$, $J$ bit errors are considered to occur for device $k$, and vice versa. 

\subsection{Key Challenges in Grant-Free NOMA}
A major challenge in realizing grant-free NOMA in Sec.~\ref{Sec:SystemA} is the limited length of the spreading sequence. This limitation is inevitable due to the finite channel coherence time and scarce radio resources available for data transmission. This challenge becomes even more crucial in scenarios involving critical or low-latency mMTC, where a small length of spreading sequences is necessary to meet stringent latency requirements.
Due to this inherent limitation in spreading sequence length, it is impossible to guarantee the optimality of existing AD, CE, and DD techniques based on the CS approach. Another key challenge comes from the lack\footnote{Typically, in grant-free NOMA, attaining the above information may require additional communication overhead between the BS and each device, which can lead to increased resource consumption and potential delays in the communication process.} of the knowledge of channel gain orders. This can lead to significant error propagation of the SIC approach for AD, CE, and DD techniques. Although existing PIC-based CE techniques do not require prior knowledge about channel gain orders, they can still suffer from significant error propagation in grant-free NOMA due to non-orthogonal spreading sequences and the absence of accurate knowledge about device activities. To establish grant-free NOMA as a viable solution, it is crucial to devise a comprehensive framework overcoming the aforementioned challenges without compromising the performance of AD, CE, and DD.
\begin{figure*}[t]
    \centering 
    {\epsfig{file=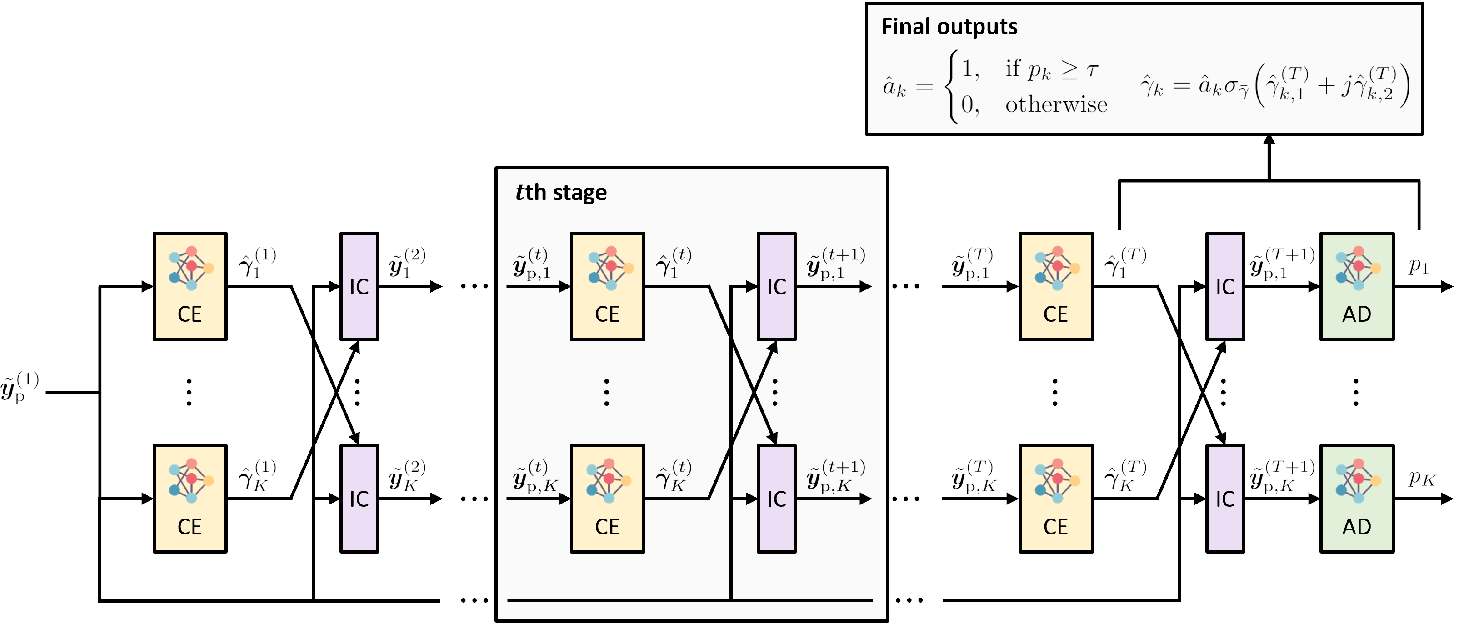,width=17cm}}
    \caption{Illustration of the pilot-only PIC framework with $T$ stages and $K$ AD modules.}  
    \label{fig:framework}
\end{figure*}

\section{Proposed Pilot-Only PIC Framework}\label{Sec:Comp}
In this section, we present a pilot-only PIC framework based on DL, which enables the accurate and efficient joint estimation of device activities and their associated channels. Our key strategy is to eliminate inter-user interference from the received pilot signal ${\bm y}_{\rm p}$ in an iterative and parallel manner. This strategy also involves addressing the CE problem with the consideration of sparse device activity patterns in grant-free NOMA, while tackling the AD problem.
In what follows, we first introduce the overall structure of the proposed framework and then present details of the training procedure. 

\subsection{Overall Structure}\label{Sec:CompA}
The proposed pilot-only PIC framework includes $T$ stages, each consisting of $K$ CE modules and $K$ IC modules. 
Following these stages, we integrate $K$ AD modules that estimate the activity of each device in a parallel manner. 
The high-level illustration of the proposed framework is presented in Fig.~\ref{fig:framework}, while detailed explanations of each component of our framework are provided below. 
\subsubsection{Input and Output}\label{SubSubSec:PIC_in_out}
The proposed framework aims to accurately estimate both the effective channel ${\gamma}_k$ and the device activity indicator $a_k$ from the received pilot signal ${\bm y}_{\rm p}$. To achieve this goal, we apply two preprocessing steps to ${\gamma}_k$ and ${\bm y}_{\rm p}$: (i) real-domain conversion and (ii) standardization.
In the real-domain conversion step, we transform the effective channel ${\gamma}_k$ and the received signal ${\bm y}_{\rm p}$, originally represented in the complex domain, into the real domain:
$\bar{\bm \gamma}_k = [\Re({\gamma}_k), \Im({\gamma}_k)]^{\sf T}\in\mathbb{R}^{2}$ and $\bar{\bm y}_{\rm p} = [\Re({\bm y}_{\rm p}^{\sf T}), \Im({\bm y}_{\rm p}^{\sf T})]^{\sf T}\in\mathbb{R}^{2L}$. 
This conversion allows us to establish a relationship between $\bar{\bm y}_{\rm p}$ and $\{\bar{\bm \gamma}_k\}_{\forall k}$, as follows:
\begin{align}\label{eq:Real}
    \bar{\bm y}_{\rm p}
    =
    \sum_{k=1}^K 
    \begin{bmatrix}
        \Re({\bm s}_k) & -\Im({\bm s}_k)\\
        \Im({\bm s}_k) & \Re({\bm s}_k)
    \end{bmatrix}
    \bar{\bm \gamma}_k
    +
    \begin{bmatrix}
    \Re({\bm n}_{\rm p}) \\
    \Im({\bm n}_{\rm p})
    \end{bmatrix}.
\end{align}
The next step involves standardization, a crucial process for improving the performance of the deep neural network (DNN).
We first compute the standard deviations $\sigma_{\bar{\gamma}}$ and $\sigma_{\bar{y}_{\rm p}}$ for the entries of $\{\bar{\bm \gamma}_k\}_{\forall k}$ and $\bar{\bm y}_{\rm p}$, respectively, utilizing the training dataset stored at the BS. 
After obtaining the standard deviations, we standardize input data as follows: 
$\tilde{\bm \gamma}_k = \bar{\bm \gamma}_k/\sigma_{\bar{\gamma}}, \forall k$ and $\tilde{\bm y}_{\rm p}^{(1)} = \bar{\bm y}_{\rm p}/\sigma_{\bar{y}_{\rm p}}$. This process yields the standardized received signal given by 
\begin{align}\label{eq:rel}
    \tilde{\bm y}_{\rm p}^{(1)}
    =
    \sum_{k=1}^K 
    \begin{bmatrix}
        \Re(\tilde{\bm s}_k) & -\Im(\tilde{\bm s}_k)\\
        \Im(\tilde{\bm s}_k) & \Re(\tilde{\bm s}_k)
    \end{bmatrix}
    \tilde{\bm \gamma}_k
    +
    \frac{1}{\sigma_{\bar{y}}}
    \begin{bmatrix}
    \Re({\bm n}_{\rm p}) \\
    \Im({\bm n}_{\rm p})
    \end{bmatrix},
\end{align}
where $\tilde{\bm s}_k =( {\sigma_{\bar{\gamma}}}/{\sigma_{\bar{y}_{\rm p}}}){\bm s}_k, \forall k$. 
Building upon the relationship in \eqref{eq:rel}, our DNN takes ${\tilde{\bm y}}_{\rm p}^{(1)}$ as input and aims to predict both $\{{\tilde{\bm \gamma}}_k\}_{\forall k}$ and $\{a_k\}_{\forall k}$ with high accuracy.

\subsubsection{Stage ($K$ CE modules and $K$ IC modules)} 
The core process of conventional PIC-based CE techniques is to eliminate inter-user interference from the received signal in an iterative and parallel manner \cite{IC_survey,DeepNOMA,PIC_CE}. This approach offers a distinct advantage over SIC-based CE techniques, as it does not require prior knowledge of channel gain orders. 
However, conventional PIC techniques are not designed for grant-free NOMA, and the use of such techniques in grant-free NOMA can introduce significant error propagation due to the use of non-orthogonal spreading sequences and the absence of accurate knowledge about device activities. 
To overcome this problem and fully harness the advantages of PIC, we incorporate DL and PIC. This integrating approach enables us to solve complex and intractable problems in grant-free NOMA, including AD, CE, and DD, based on a data-driven manner. Furthermore, we can improve both performance and computational efficiency by integrating communication knowledge related to PIC into DL, rather than treating DL as a black box. 
Given that our framework consists of a sequence of $T$ uniform stages, the subsequent sections in this part will focus on elucidating the operations in stage $t\in\{1,\ldots,T\}$. 

Let ${\bm \theta}_{\rm CE}^{(t,k)}$ be the parameter vector of the $k$th CE module in  stage $t$. 
The mapping function of the CE module is defined as
\begin{align}
    \hat{\bm \gamma}_k^{(t)} = f_{{\bm \theta}_{\rm CE}^{(t,k)}}(\tilde{\bm y}_{{\rm p}, k}^{(t)}).  
\end{align}
Here, $\tilde{\bm y}_{{\rm p}, k}^{(t)}$ represents an output of the IC module, defined as
\begin{align}\label{eq:IC}
    \tilde{\bm y}_{{\rm p},k}^{(t)} = \tilde{\bm y}_{\rm p}^{(1)} - \sum_{i=1, i \neq k}^K 
    \begin{bmatrix}
        \Re(\tilde{\bm s}_i) & -\Im(\tilde{\bm s}_i)\\
        \Im(\tilde{\bm s}_i) & \Re(\tilde{\bm s}_i)
    \end{bmatrix}
    \hat{\bm \gamma}_i^{(t-1)},
\end{align}
where $\hat{\bm \gamma}_i^{(0)} = {\bm 0}_2, \forall i\in\{1,\ldots,K\}$. 
In our IC module, interferences with estimated channels in the previous stage are removed from the superimposed received signal in \eqref{eq:rel}. This interference removal process is carried out in parallel, similar to conventional PIC, and is essential for improving CE accuracy. It is noteworthy that all CE modules in our framework allow back-propagation of the gradients and, therefore, can be jointly trained in an end-to-end manner. 



\subsubsection{AD Module} 
In the AD module, we employ a DNN, which is individually parameterized by ${\bm \theta}_{\rm AD}^{(k)}$ for each device $k$. Each AD module, associated with device $k$, estimates the device activity indicator $a_k$ from the received signal after interference removal processes across $T$ stages. This design facilitates efficient AD without the need for an excessively complex or over-parameterized model.


For the final layer of all $K$ AD modules, we utilize the sigmoid activation function. This decision makes our training process smoother as we use true device activity indicators represented by binary labels.
The mapping function for each AD module is represented as follows:
\begin{align}
    p_k = f_{{\bm \theta}_{\rm AD}^{(k)}}(\tilde{\bm y}_{{\rm p}, k}^{(T+1)}),
\end{align}
where $p_k$ is the activity score within the range of $(0,1)$ and $\tilde{\bm y}_{{\rm p}, k}^{(T+1)}$ is the output from the last stage of the proposed framework. 
It should be noted that $\tilde{\bm y}_{{\rm p}, k}^{(T+1)}$ represents the received signal after interference removal processes during the previous $T$ stages. This fact implies that the device activity can be more accurately estimated from $\tilde{\bm y}_{{\rm p}, k}^{(T+1)}$ than simply using the original received signal $\tilde{\bm y}_{\rm p}^{(1)}$ without interference removal.



\subsubsection{Final Decision} 
In the grant-free NOMA system, the activity indicator should be either zero or one. To do so, we perform the thresholding of $p_k$ to generate the activity indicator $a_k$ for device $k$. That is, 
\begin{align}\label{eq:est_act}
    \hat{a}_k = 
    \begin{cases} 
        {1,}&{\text{if }p_k\geq \tau},\\ 
        {0,}&{\text{otherwise}}, 
    \end{cases}
\end{align}
where $\tau$ is a pre-defined threshold that balances the trade-off between the false-alarm and miss-detection probabilities. It should be noted we treat $\tau$ as a hyper-parameter to provide flexibility applicable to various grant-free NOMA scenarios. 
Meanwhile, the estimated channel should take on zero value when the device is determined to be inactive. To ensure this, we set an estimate of the channel based on the estimated activity indicator in \eqref{eq:est_act} and the standardization in \eqref{eq:rel} as follows:
\begin{align}\label{eq:est_ch}
    \hat{\gamma}_k = \hat{a}_k\sigma_{\bar{\gamma}} \Big(\hat{\gamma}_{k,1}^{(T)} + j\hat{\gamma}_{k,2}^{(T)}\Big),
\end{align}
where $\hat{\gamma}_{k,i}^{(T)}$ is the $i$th component of $\hat{\bm \gamma}_k^{(T)}$. After obtaining the estimated effective channel in \eqref{eq:est_ch}, various detection methods, including minimum mean squared error, SIC, or maximal-ratio combining (MRC), can be employed to detect the symbols $\{x_{k,d}\}_{\forall k,d}$. In this work, we focus on the MRC \cite{CS1} for simplicity:  
\begin{align}\label{eq:est_data}
     \bar{{\bm y}}_{{\rm d},d,k}&= \frac{{\bm s}_k^{\sf H}}{{\hat{\gamma}_k}}{\bm y}_{{\rm d},d} \nonumber\\&=  \frac{{\gamma}_k}{\hat{\gamma}_k}x_{k,d} + \frac{1}{{\hat{\gamma}_k}}\sum_{i=1,i\neq k}^K {\bm s}_k^{\sf H}{\bm s}_i\gamma_ix_{i,d} + \frac{{\bm s}_k^{\sf H}}{{\hat{\gamma}_k}}{\bm n}_{{\rm d},d},
\end{align}
Note that MRC is applied to the devices with $\hat{a}_k = 1$. Subsequently, $\hat{x}_{k,d}$ is determined from $\bar{{\bm y}}_{{\rm d},d,k}$ using the maximum likelihood detector.



\subsection{Training Procedure}\label{SubSec:PIC_train}
In the grant-free NOMA, the CE and AD problems are categorized as regression and classification problems. These two tasks are closely related, as both tasks rely on the same received signal and have dependencies stemming from activity indicators. To accurately estimate both $\{\tilde{\bm \gamma}_k\}_{\forall k}$ and $\{a_k\}_{\forall k}$, we employ a joint loss function that can optimize the performance of both regression and classification tasks.


Firstly, we use the regression loss function to measure the discrepancy between the estimated and true channels. 
\begin{align}\label{eq:loss_reg}
    \mathcal{L}_{\rm Reg} = \frac{1}{2K}\sum_{t=1}^T w_t \sum_{k=1}^K\|\hat{\bm \gamma}_k^{(t)} - \tilde{\bm \gamma}_k\|_1,
\end{align}
where $w_t$ is the weight associated with each stage. In the regression loss in \eqref{eq:loss_reg}, we leverage the mean absolute error (MAE) instead of the more commonly used mean squared error (MSE) in regression tasks \cite{Loss_MSE1,Loss_MSE2}.
Our motivation behind this choice comes from the sparse characteristic of effective channels $\{\tilde{\bm \gamma}_k\}_{\forall k}$ in grant-free NOMA systems. Specifically, this characteristic implies that while the majority of devices exhibit zero channel gains, a small subset has much higher channel gains, resulting in outliers within the dataset. 
Traditionally, the MSE loss serves as a prevalent metric used to evaluate the performance of CE \cite{Loss_MSE1,Loss_MSE2}. However, in the presence of outliers in effective channels, the MSE loss may prioritize minimizing errors, particularly for these outliers, while potentially neglecting the majority of the data \cite{boyd2004convex}. 
In contrast, MAE is known to be robust to outliers \cite{boyd2004convex}, aligning well with the characteristics of grant-free NOMA scenarios. Therefore, our decision to use MAE enhances the overall performance and robustness of our framework while mitigating error propagation.


Next, we present the classification loss function which measures the difference between the activity scores and the true activity indicators as follows:
\begin{align}\label{eq:loss_class}
    \mathcal{L}_{\rm Class} = -\frac{1}{K}\sum_{k=1}^K\left[a_k\log p_k + (1-a_k)\log(1-p_k)\right].
\end{align}
This loss function is commonly known as the binary cross-entropy (BCE) loss function and is highly effective for binary label training. By adopting the BCE loss function, we can effectively train $K$ AD modules as well as the entire framework.
Combining the loss functions in \eqref{eq:loss_reg} and \eqref{eq:loss_class}, we obtain the joint loss function: 
\begin{align}\label{eq:loss_all}
   \mathcal{L} = \lambda \mathcal{L}_{\rm Class} + (1-\lambda)\mathcal{L}_{\rm Reg},
\end{align}
where $\lambda\in(0,1)$ is a hyper-parameter determined by the relative importance of CE performance compared to AD performance. 


Our proposed framework offers end-to-end training facilitated by the IC module and joint loss function \eqref{eq:loss_all}, enhancing performance by capturing complex inter-module relationships. Our framework also addresses the vanishing gradient problem by computing MAE loss at each stage $t$, not just the final stage. Additionally, our framework integrates joint training for regression (CE) and classification (AD) tasks. This feature not only streamlines the training pipeline but also reduces the size of the DNN compared to strategies that separate regression and classification tasks.


\section{Extension of the Pilot-Only PIC}\label{Sec:Extend}
In this section, we introduce two promising extensions of the pilot-only PIC framework. 
The first extension, referred to as {\em data-aided PIC}, is designed to extract common channel information shared between the received pilot signal ${\bm y}_{\rm p}$ and the received data signals $\{{\bm y}_{{\rm d}, d}\}_{\forall d}$. By jointly utilizing received pilot and data signals, this framework significantly enhances the overall performance in coherent scheme.  
The second extension, referred to as non-coherent PIC, is designed to support non-coherent scheme in grant-free NOMA scenarios, transmitting a small number of data bits without spreading sequences. Leveraging the benefits of non-coherent scheme, this framework provides a valuable solution for enhancing communication efficiency in grant-free NOMA, particularly when the number of data bits to be transmitted is small.
In the remainder of this section, we first introduce the data-aided PIC framework and then present the non-coherent PIC framework. 

\begin{figure}[t]
    \centering 
    {\epsfig{file=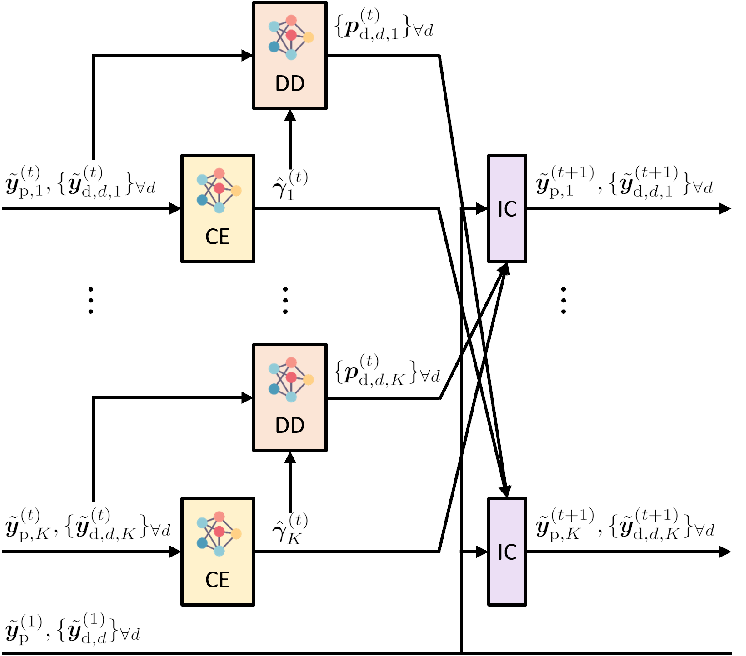,width=8cm}}
    \caption{Illustration of the $t$th stage in the data-aided PIC framework, consisting of $K$ CE modules, $K$ IC modules, and $K$ DD modules.}
    \label{fig:PIC_da}
\end{figure}

\subsection{Data-Aided PIC} The proposed data-aided PIC framework includes $T$ uniform stages, each consisting of $K$ CE modules, $K$ IC modules, and $K$ DD modules. A notable distinction between the data-aided PIC framework and the pilot-only PIC framework lies in the inclusion of the DD module and the exclusion of the AD module. The DD module is responsible for estimating both the device activity indicator $a_k$ and the data symbols $\{x_{k,d}\}_{\forall d}$. Another notable distinction is the input of the CE module, which includes $\{{\bm y}_{{\rm d},d}\}_{\forall d}$ as well as ${\bm y}_{\rm p}$ to extract the common channel information between received signals. 
The high-level illustration of the $t$th stage of the data-aided PIC framework is depicted in Fig.~\ref{fig:PIC_da}, with detailed explanations provided below. 

\subsubsection{Input and Output} The proposed data-aided PIC framework aims to accurately estimate the effective channels $\{\gamma_k\}_{\forall k}$, device activity indicators $\{a_k\}_{\forall k}$, and data symbols $\{x_{k,d}\}_{\forall k,d}$ from the received signals ${\bm y}_{\rm p}$ and $\{{\bm y}_{{\rm d},d}\}_{\forall d}$. 
To achieve this goal, real-domain conversion and standardization are applied to $\{\gamma_k\}_{\forall k}$, ${\bm y}_{\rm p}$, and $\{{\bm y}_{{\rm d},d}\}_{\forall d}$, as described in Sec.~\ref{Sec:CompA}. The resulting effective channel and received signals are denoted as $\{\tilde{{\bm \gamma}}_k\}_{\forall k}$, $\tilde{\bm y}_{\rm p}^{(1)}$, and $\{\tilde{\bm y}_{{\rm d},d}^{(1)}\}_{\forall d}$. Our DNN takes $\tilde{\bm y}_{\rm p}^{(1)}$ and $\{\tilde{\bm y}_{{\rm d},d}^{(1)}\}_{\forall d}$ as input and aims to predict $\{\tilde{{\bm \gamma}}_k\}_{\forall k}$, $\{a_k\}_{\forall k}$, and $\{x_{k,d}\}_{\forall k,d}$.

\subsubsection{Stage ($K$ CE modules, $K$ IC modules, and $K$ DD modules)} 
In each stage of the data-aided PIC framework, the effective channel $\tilde{\bm \gamma}_k$ is initially estimated, followed by the estimation of data symbols $\{x_{k,d}\}_{\forall d}$ based on the obtained effective channel. After obtaining the estimated channel and data symbols, the PIC is performed. 
Let ${\bm \theta}_{\rm CE}^{(t,k)}$ be the parameter vector of the $k$th CE module in stage $t$. The mapping function of the CE module is defined as
\begin{align}\label{eq:CE_DA}
    \hat{\bm \gamma}_k^{(t)} = f_{{\bm \theta}_{\rm CE}^{(t,k)}}(\tilde{\bm y}_{{\rm p}, k}^{(t)}, \{\tilde{\bm y}_{{\rm d}, d, k}^{(t)}\}_{\forall d}),
\end{align}
where $\tilde{\bm y}_{{\rm p}, k}^{(t)}$ and $\tilde{\bm y}_{{\rm d}, d, k}^{(t)}$ represent the outputs of the $k$th IC module in stage $(t-1)$ whose definitions will be provided later. Let ${\bm \theta}_{\rm DD}^{(t,k)}$ be the parameter vector of the $k$th DD module in stage $t$. For the last layer of all DD modules, the sigmoid activation function is utilized. 
The mapping function of the DD module is represented as
\begin{align}
    \{{\bm p}_{{\rm d}, d, k}^{(t)}\}_{\forall d} = f_{{\bm \theta}_{\rm DD}^{(t,k)}}(\{\tilde{\bm y}_{{\rm d}, d, k}^{(t)}\}_{\forall d}, \hat{\bm \gamma}_k^{(t)}),  
\end{align}
where ${\bm p}_{{\rm d}, d, k}^{(t)} = [{\bm p}_{{\rm d}, d, k,1}^{(t)},\ldots,{\bm p}_{{\rm d}, d, k,|\mathcal{C}|}^{(t)}]^{\sf T}\in (0,1)^{|\mathcal{C}|}$ and ${\bm p}_{{\rm d}, d, k,c}^{(t)}$ implies the probability that $x_{k,d}$ is the $c$th symbol in the constellation $\mathcal{C}$. Given $\{{\bm p}_{{\rm d}, d, k}^{(t)}\}_{\forall d}$ and $\hat{\bm \gamma}_k^{(t)}$, the IC module initiates the following PIC processes:
\begin{align}
    \tilde{\bm y}_{{\rm p},k}^{(t+1)} &= \tilde{\bm y}_{\rm p}^{(1)} - \sum_{i=1, i \neq k}^K 
    \begin{bmatrix}
        \Re(\tilde{\bm s}_i) & -\Im(\tilde{\bm s}_i)\\
        \Im(\tilde{\bm s}_i) & \Re(\tilde{\bm s}_i)
    \end{bmatrix}
    \hat{\bm \gamma}_i^{(t)},
\end{align}
and
\begin{align}\label{eq:PIC_data}
    \tilde{\bm y}_{{\rm d},d,k}^{(t+1)} &= \tilde{\bm y}_{{\rm d},d}^{(1)} - \Bigg\{\sum_{i=1, i \neq k}^K 
    \begin{bmatrix}
        \Re(\tilde{\bm s}_i) & -\Im(\tilde{\bm s}_i)\\
        \Im(\tilde{\bm s}_i) & \Re(\tilde{\bm s}_i)
    \end{bmatrix} \nonumber \\
    &~~~
    \times
    \begin{bmatrix}
        \hat{\gamma}_{i,1}^{(t)} & -\hat{\gamma}_{i,2}^{(t)}\\
        \hat{\gamma}_{i,2}^{(t)} & \hat{\gamma}_{i,1}^{(t)} 
    \end{bmatrix} 
    \begin{bmatrix}
        \Re({\bm x}_{\mathcal{C}}^{\sf T}{\bm p}_{{\rm d},d,k}^{(t)}) \\
        \Im({\bm x}_{\mathcal{C}}^{\sf T}{\bm p}_{{\rm d},d,k}^{(t)})
    \end{bmatrix}\Bigg\},
\end{align}
where ${\bm x}_{\mathcal{C}}$ is a symbol vector comprising all symbols in $\mathcal{C}$. If the constellation set is binary phase-shift keying (BPSK), i.e., $\mathcal{C} = \{-1,1\}$, the PIC process in \eqref{eq:PIC_data} can be simplified as follows:
\begin{align}
    \tilde{\bm y}_{{\rm d},d,k}^{(t+1)} &= \tilde{\bm y}_{{\rm d},d}^{(1)} - \sum_{i=1, i \neq k}^K 
    \begin{bmatrix}
        \Re(\tilde{\bm s}_i) & -\Im(\tilde{\bm s}_i)\\
        \Im(\tilde{\bm s}_i) & \Re(\tilde{\bm s}_i)
    \end{bmatrix} \nonumber \\
    &~~~\times\hat{\bm \gamma}_i^{(t)}(p_{{\rm d},d,k,1}^{(t)} - p_{{\rm d},d,k,2}^{(t)}),
\end{align}
where $p_{{\rm d},d,k,1}^{(t)}$ and $p_{{\rm d},d,k,2}^{(t)}$ represent the first and second components of ${\bm p}_{{\rm d},d,k}^{(t)}$, respectively, and have the meanings of $\mathbb{P}[x_{k,d} = 1]$ and $\mathbb{P}[x_{k,d} = -1]$, respectively. 

It should be noted that, in the design of the CE module, various DNN structures, including a convolutional neural network, can be considered to effectively extract common channel information from the received pilot and data signals. 
In our framework, we use the fully connected neural network (FCNN) whose input is structured as $[(\tilde{\bm y}_{{\rm p},k}^{(t)})^{\sf T}, (\tilde{\bm y}_{{\rm d},1,k}^{(t)})^{\sf T}, \ldots, (\tilde{\bm y}_{{\rm d},D,k}^{(t)})^{\sf T}]^{\sf T}$ in order to reduce the overall complexity. 
Similarly, the DD module also adopts an FCNN with its input structured as $[(\tilde{\bm y}_{{\rm d},1,k}^{(t)})^{\sf T}, \ldots, (\tilde{\bm y}_{{\rm d},D,k}^{(t)})^{\sf T}, (\hat{\bm \gamma}_k^{(t)})^{\sf T}]^{\sf T}$ and its output organized as $[({\bm p}_{{\rm d},1,k}^{(t)})^{\sf T}, \ldots, ({\bm p}_{{\rm d},D,k}^{(t)})^{\sf T}]^{\sf T}$.



\subsubsection{Final Decision}
The final procedure of the data-aided PIC framework is to estimate the device activity indicators $\{a_k\}_{\forall k}$ and the data bits $\{b_{k,d}\}_{\forall k,j}$. 
To accomplish this, we first compute the maximum value of ${\bm p}_{{\rm d},d,k}^{(T)}$ as
\begin{align}
    {p}_{{\rm max},d,k} = \max\!\Big\{{p}_{{\rm d},d,k,1}^{(T)}, \ldots, {p}_{{\rm d},d,k,|\mathcal{C}|}^{(T)}\Big\},
\end{align}
implying the maximum probability that a certain symbol in $\mathcal{C}$ is transmitted. Indeed, $p_{{\rm max},d,k}$ should be zero if the device does not transmit any symbol, i.e., the device is inactive. Our data-aided PIC framework can address this circumstance because the DD module with the sigmoid function can naturally produce low values of $\{{p}_{{\rm d},d,k,c}^{(T)}\}_{\forall c}$ if the symbol is not transmitted. 
After obtaining $\{{p}_{{\rm max},d,k}\}_{\forall d}$, we compute their mean as
\begin{align}
    \mu_{k} = \frac{1}{D}\sum_{d=1}^D p_{{\rm max},d,k} \in (0,1),
\end{align}
which quantifies the probability of device $k$ being activated and transmitting $D$ symbols. Subsequently, we determine an estimate of the activity indicator for device $k$ as 
\begin{align}
    \hat{a}_k = 
    \begin{cases} 
        {1,}&{\text{if }\mu_k\geq \tau},\\ 
        {0,}&{\text{otherwise}},
    \end{cases}
\end{align}
We note that the data-aided PIC framework does not explicitly utilize the AD module, a component leveraged in the pilot-only PIC framework in Sec.~\ref{Sec:Comp}. This strategic choice not only reduces the complexity of the data-aided framework but also enhances its training efficiency compared to strategies that separate AD and DD tasks. 
After obtaining $\hat{a}_k$, we finally determine an estimate of the symbol $x_{k,d}$ as 
\begin{align}\label{eq:est_data_DA}
    \hat{x}_{k,d} = 
    \begin{cases} 
        {x_{\mathcal{C},c},}&{\text{if }\hat{a}_k = 1\text{ and }p_{{\rm max},d,k}=p_{{\rm d},d,k,c}^{(T)}},\\
        {0,}&{\text{otherwise}},
    \end{cases}
\end{align}
where ${x_{\mathcal{C},c}}$ is the $c$th component of ${\bm x}_{\mathcal{C}}$. Here, the second line in \eqref{eq:est_data_DA} implies device $k$ is estimated to be inactive. Following this, an estimate of the data bit is readily determined through the demodulation process. 
Meanwhile, to evaluate the CE performance of the data-aided PIC framework, we can determine an estimate of the channel based on $\hat{a}_k$, $\hat{\bm \gamma}_k^{(T)}$, and the equation in \eqref{eq:est_ch}. 

\subsubsection{Training Procedure} 
The overall training strategy of the data-aided PIC framework is similar to that of the pilot-only PIC framework. We first categorize the CE and DD problems as regression and classification problems. These two tasks are closely related, as both tasks rely on the same received data signals and have dependencies stemming from activity indicators. To address this interdependence, we employ a joint loss function that can optimize the performance of both regression and classification tasks. This comprehensive training approach facilitates the effective learning and coordination of the two tasks within the data-aided PIC framework.




Firstly, we utilize the regression loss function, quantifying the disparity between the estimated channels and the true channels, as the MAE-based loss function in \eqref{eq:loss_reg}. Further details regarding the rationale for adopting this function are elaborated in Sec.~\ref{SubSec:PIC_train}. By modifying the loss function in \eqref{eq:loss_class}, we also present the classification loss function to measure the performance of data detection, as follows: 
\begin{align}\label{eq:cl_loss_da}
    \mathcal{L}_{\rm ClassDA} = \frac{1}{KD|\mathcal{C}|}\sum_{t=1}^T\sum_{k=1}^K\sum_{d=1}^D\sum_{c=1}^{|\mathcal{C}|}\mathcal{L}_{\rm ClassDA}^{(t,d,k,c)},
\end{align}
and 
\begin{align}
    \mathcal{L}_{\rm ClassDA}^{(t,d,k,c)}\!=\! -e_{k,d,c}\log p_{{\rm d},d,k,c}^{(t)} - (1-e_{k,d,c})\log(1-p_{{\rm d},d,k,c}^{(t)}),
\end{align}
where $e_{k,d,c}\in\{0,1\}$ indicates whether or not device $k$ transmits the $c$th symbol in $\mathcal{C}$ at the $d$th data transmission time. 
Similar to \eqref{eq:loss_all}, combining the loss functions in \eqref{eq:loss_reg} and \eqref{eq:cl_loss_da}, we obtain the joint loss function for training the data-aided PIC framework. By minimizing this joint loss function, collaborative training for all CE and DD modules can be achieved. 

\begin{figure}[t]
    \centering 
    {\epsfig{file=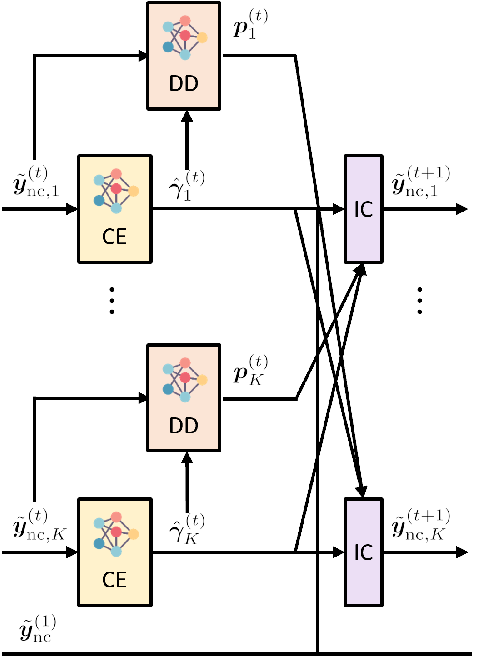,width=5.5cm}}
    \caption{Illustration of the $t$th stage in the non-coherent PIC framework, consisting of $K$ CE modules, $K$ IC modules, and $K$ DD modules.}  
    \label{fig:PIC_nch}
\end{figure}

\subsection{Non-Coherent PIC}
The proposed non-coherent PIC framework includes $T$ uniform stages, each consisting of $K$ CE modules, $K$ IC modules, and $K$ DD modules. Unlike the pilot-only and data-aided frameworks, this framework is specifically designed for the non-coherent scheme introduced in Sec.~\ref{Sec:SystemA}. The high-level illustration of the $t$th stage of the non-coherent PIC framework is depicted in Fig.~\ref{fig:PIC_nch}. 

\subsubsection{Input and Output} The proposed non-coherent PIC framework aims to accurately estimate the $\{\gamma_k\}_{\forall k}$, $\{a_{k,j}\}_{\forall k,j}$ and $\{a_{k}\}_{\forall k}$.
To achieve this goal, we apply real-domain conversion and standardization to $\{\gamma_{k,j}\}_{\forall k,j}$ and ${\bm y}_{\rm nc}$, as described in Sec.~\ref{Sec:CompA}. The resulting effective channel and received signals are denoted as $\{\tilde{{\bm \gamma}}_{k,j}\}_{\forall k,j}$ and $\tilde{\bm y}_{\rm nc}^{(1)}$, respectively. Our DNN takes $\tilde{\bm y}_{\rm nc}^{(1)}$ as input and aims to predict $\{\tilde{{\bm \gamma}}_k\}_{\forall k}$, $\{a_{k,j}\}_{\forall k,j}$, and $\{a_k\}_{\forall k}$. 

\subsubsection{Stage ($K$ CE modules, $K$ IC modules, and $K$ DD modules)}
In each stage of the non-coherent PIC framework, the effective channel $\tilde{\bm \gamma}_{k}\triangleq\sum_{j=1}^{2^J}\tilde{\bm \gamma}_{k,j}$ is initially estimated, followed by the estimation of $\{a_{k,j}\}_{\forall k}$ based on the obtained effective channel. Subsequently, the PIC is performed using the IC module. 
Let ${\bm \theta}_{\rm CE}^{(t,k)}$ be the parameter vector of the $k$th CE module in stage $t$. The mapping function of the CE module is defined as
\begin{align}\label{eq:CE_NC}
    \hat{\bm \gamma}_k^{(t)} \triangleq \sum_{j=1}^{2^J}\hat{\bm \gamma}_{k,j}^{(t)} = f_{{\bm \theta}_{\rm CE}^{(t,k)}}(\tilde{\bm y}_{{\rm nc},k}^{(t)}),
\end{align}
where $\tilde{\bm y}_{{\rm nc},k}^{(t)}$ is the outputs of the $k$th IC module in stage $(t-1)$, and its definition will be provided later. Note that the true effective channel $\tilde{\bm \gamma}_{k}=\sum_{j=1}^{2^J}\tilde{\bm \gamma}_{k,j}$ becomes ${\bm 0}_2$ when the device $k$ is inactive. Conversely, the true effective channel takes on a non-zero value $\tilde{\bm \gamma}_{k,i}$ when the device $k$ is active with $a_{k,i}=1$. 
In this regard, the CE module in \eqref{eq:CE_NC} estimates the sum of $\hat{\bm \gamma}_{k,1}^{(t)},\ldots,\hat{\bm \gamma}_{k,2^J}^{(t)}$, rather than individually estimating each $\hat{\bm \gamma}_{k,j}^{(t)}$. This strategy effectively reduces the overall complexity of the proposed framework, while facilitating the training process. 
Next, let ${\bm \theta}_{\rm DD}^{(t,k)}$ be the parameter vector of the $k$th DD module in stage $t$. For the last layer of all DD modules, the sigmoid activation function is utilized. 
The mapping function of the DD module is represented as follows: 
\begin{align}
    {\bm p}_{k}^{(t)} = f_{{\bm \theta}_{\rm DD}^{(t,k)}}(\tilde{\bm y}_{{\rm nc},k}^{(t)}, \hat{\bm \gamma}_k^{(t)}),  
\end{align}
where ${\bm p}_{k}^{(t)} = [p_{k,1}^{(t)},\ldots,p_{k,2^J}^{(t)}]^{\sf T}\in(0,1)^{2^J}$ and $p_{k,j}^{(t)}$ implies the probability that $a_{k,j} = 1$. Subsequently, the IC module performs the following PIC processes:
\begin{align}
    \tilde{\bm y}_{{\rm nc},k}^{(t+1)} &= \tilde{\bm y}_{{\rm nc}}^{(1)} - \sum_{i=1, i \neq k}^K \sum_{j=1}^{2^J}
    \begin{bmatrix}
        \Re(\tilde{\bm s}_i) & -\Im(\tilde{\bm s}_i)\\
        \Im(\tilde{\bm s}_i) & \Re(\tilde{\bm s}_i)
    \end{bmatrix}
    \hat{\bm \gamma}_i^{(t)}p_{i,j}^{(t)},
\end{align}
where $\hat{\bm \gamma}_i^{(t)}p_{i,j}^{(t)}$ represents the estimate of $\hat{\bm \gamma}_{i,j}^{(t)}, i\in\{1,\ldots,K\}$. Aligned with the design principle of the data-aided PIC framework, the DD module of the non-coherent PIC framework is configured as an FCNN, with its input structured as $[(\tilde{\bm y}_{{\rm nc},k}^{(t)})^{\sf T}, (\hat{\bm \gamma}_k^{(t)})^{\sf T}]^{\sf T}$. 

\subsubsection{Final Decision} 
The final procedure of the non-coherent PIC framework is to estimate $\{a_k\}_{\forall k}$ and $\{a_{k,j}\}_{\forall k,j}$. 
To do so, we first compute the maximum value of ${\bm p}_{k}^{(T)}$ as
\begin{align}
    {p}_{{\rm max},k} = \max\!\Big\{{p}_{k,1}^{(T)}, \ldots, {p}_{k,2^J}^{(T)}\Big\}.
\end{align}
Subsequently, we determine an estimate of the activity indicator for device $k$ as 
\begin{align}
    \hat{a}_k = 
    \begin{cases} 
        {1,}&{\text{if }{p}_{{\rm max},k}\geq \tau},\\ 
        {0,}&{\text{otherwise}},
    \end{cases}
\end{align}
where $\tau$ is the pre-defined threshold discussed in Sec.~\ref{Sec:Comp}.
We finally determine an estimate of $a_{k,j}$ as 
\begin{align}\label{eq:akj_NC}
    \hat{a}_{k,j} = 
    \begin{cases} 
        {1,}&{\text{if }\hat{a}_k = 1\text{ and }p_{{\rm max},k}=p_{k,j}^{(T)}},\\
        {0,}&{\text{otherwise}}.
    \end{cases}
\end{align}
It should be noted that the decision procedure in \eqref{eq:akj_NC} satisfies 
\begin{align}
    \sum_{j=1}^{2^J}\hat{a}_{k,j} = 
    \begin{cases} 
        {1,}&{\text{if } \hat{a}_k = 1},\\ 
        {0,}&{\text{if } \hat{a}_k = 0}.
    \end{cases}
\end{align}
This property involves the unique characteristic in the non-coherent scheme, given in \eqref{eq:true_akj_NC}, wherein at most one $\hat{a}_{k,j}$ is equal to one if the device $k$ is determined to be active. 
\begin{figure*}[t]
    \centering 
    \subfigure[$\epsilon = 0.1$]
    {\epsfig{file=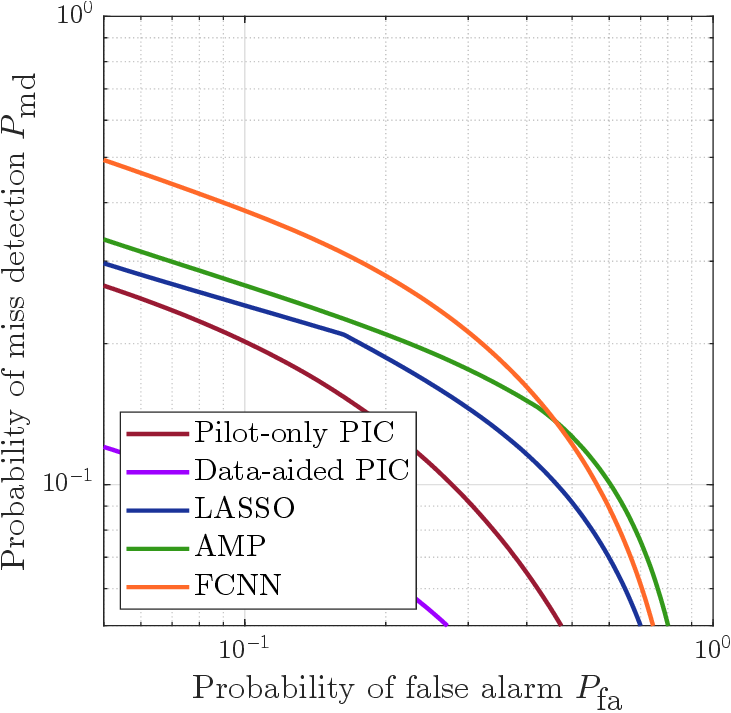, width=4.5cm}}
    \subfigure[$\epsilon = 0.2$]
    {\epsfig{file=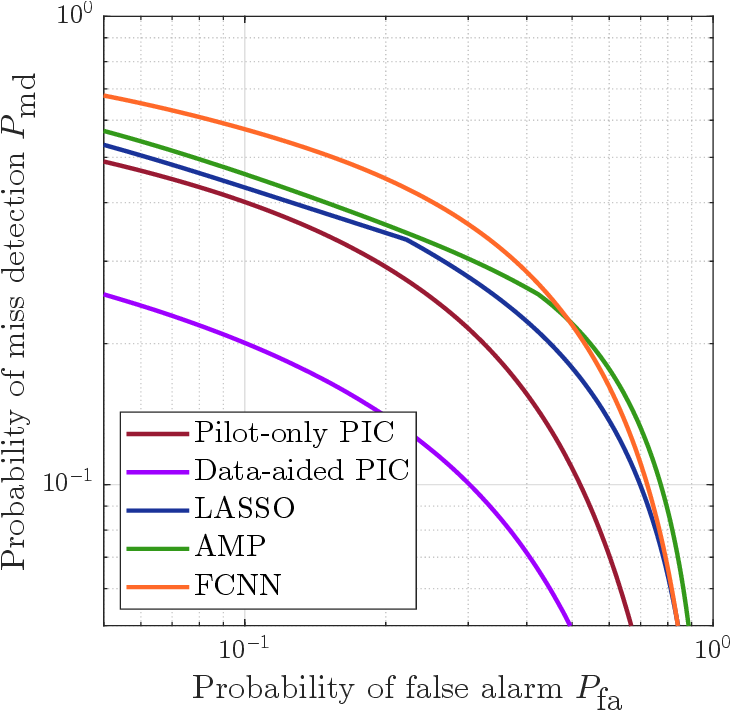, width=4.5cm}}
    \subfigure[$\epsilon = 0.3$]
    {\epsfig{file=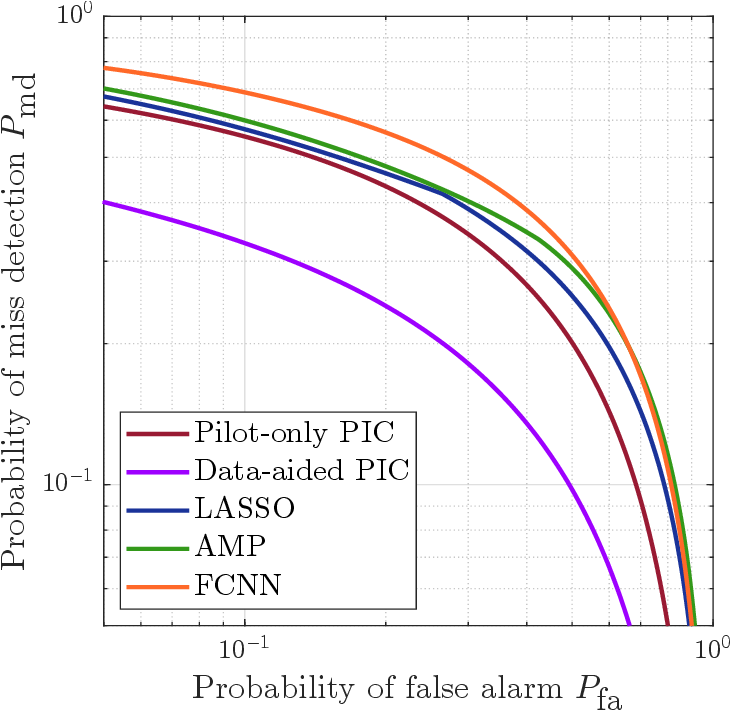, width=4.5cm}}
    \subfigure[$\epsilon = 0.4$]
    {\epsfig{file=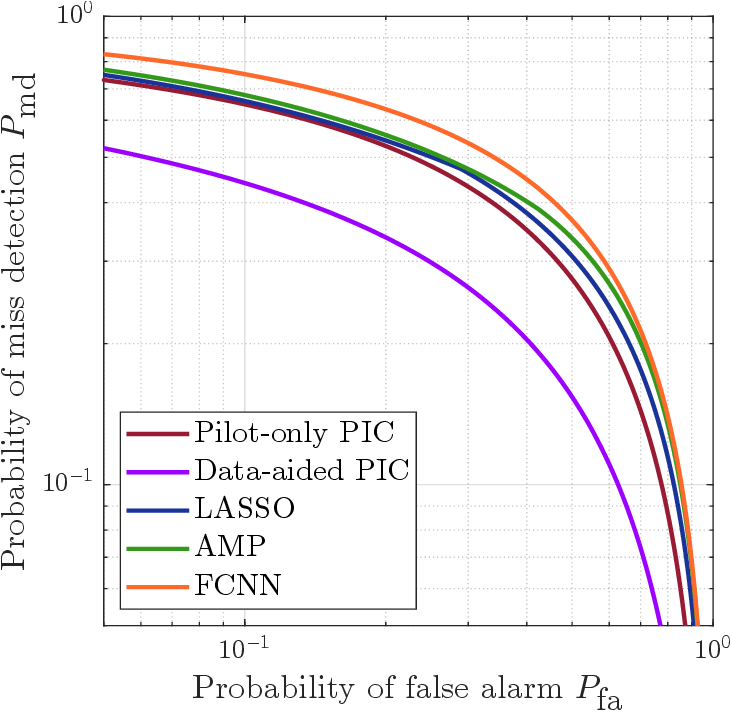, width=4.5cm}}
        \caption{Comparison of miss detection and false alarm probabilities of different grant-free NOMA frameworks with the coherent scheme for various values of $\epsilon$ when $L=6$.}  
    \label{fig:P_MD_P_FA}
\end{figure*}

\subsubsection{Training Procedure} The overall training strategy of the non-coherent PIC framework is similar to that of the pilot-only and data-aided PIC frameworks. 
We employ the regression loss function, which quantifies the disparity between the estimated channels and the true channels, as the MAE-based loss function in \eqref{eq:loss_reg}. Subsequently, we introduce the classification loss function, assessing the performance of $\{{\bm p}_{k}^{(t)}\}_{\forall k,t}$, given by 
\begin{align}\label{eq:cl_loss_nch}
    \mathcal{L}_{\rm ClassNC} = \frac{1}{K2^J}\sum_{t=1}^T\sum_{k=1}^K\sum_{j=1}^{2^J}\mathcal{L}_{\rm ClassNC}^{(t,k,j)},
\end{align}
where 
\begin{align}
    \mathcal{L}_{\rm ClassNC}^{(t,k,j)} = -a_{k,j}\log p_{k,j}^{(t)} - (1-a_{k,j})\log(1-p_{k,j}^{(t)}).
\end{align}
Similar to \eqref{eq:loss_all}, the combination of the loss functions in \eqref{eq:loss_reg} and \eqref{eq:cl_loss_nch} results in the joint loss function used for training the non-coherent PIC framework. By minimizing this joint loss function, collaborative training for all CE and DD modules is achieved, leading to improved performance in CE, DD, and AD.

\section{Simulation Results and Analysis}\label{Sec:Simul}
In this section, we demonstrate the superiority of the proposed frameworks over existing AD, CE, and DD frameworks for the uplink grant-free NOMA system, using simulations. 

\subsection{Simulation Settings}
In our simulations, we consider an uplink grant-free NOMA system where $20$ devices are distributed uniformly within a cell, with their locations ranging from $50$ m to $500$ m from the center of the cell\footnote{In mMTC scenarios with an extremely large number of devices, to efficiently accommodate these devices, we can consider the strategies to form groups for these devices and allocate different communication resources to each group, as discussed in \cite{Group_BCJ}. This strategy is especially effective in critical or low-latency mMTC scenarios where the sequence length $L$ should be sufficiently small \cite{CmMTC1, JCH_PJH}.
}. 
The devices transmit spreading sequences with a power of $\rho=20$ dBm and a bandwidth of $100$ kHz \cite{Sim_BW}. For data transmission, the devices modulate the data bits using BPSK. 
The power spectral density of the AWGN at the BS is $-169$ dBm/Hz, and $\beta_k$ follows the path-loss model $-128.1-36.7\log_{10} d_k$ dB, where $d_k$ is the distance between device $k$ and the BS \cite{Sim_set,Model_driven}. We generate each spreading sequence ${\bm s}_k$ or ${\bm s}_{k,j}$ from a Gaussian distribution $\mathcal{CN}({\bm 0},\frac{1}{L}{\bm I})$ and subsequently normalize it to ensure $\|{\bm s}_k\|_2 = 1$ or $\|{\bm s}_{k,j}\|_2 = 1$. It should be noted that, unlike CS algorithms, the proposed frameworks do not impose any specific structure constraints on the spreading sequences.

For training, we first generate training data samples with $\epsilon=0.25$. Then, we perform post-processing on both the training and test datasets to ensure that every data sample contains at least one active device, which is essential for evaluating NMSE in \eqref{eq:CE}.
We train our model with a batch size of $1024$ for $100$ epochs using the Adam optimizer in \cite{ADAM} with a learning rate of $0.001$, $\lambda = 0.5$, and $w_t = 1, \forall t$.
All the AD, CE, and DD modules in our frameworks consist of two hidden layers, each comprising $64$ neurons and rectified linear unit (ReLU) activation functions. 

For comparison, in the coherent scheme, we consider the approximate message passing (AMP)  in \cite{AMP_sim}, the least absolute shrinkage and selection operator (LASSO) in \cite{LASSO}, and an FCNN. In the LASSO framework, the objective is to minimize the following loss function:
\begin{align}\label{eq:loss_LASSO}
    \mathcal{L}_{\rm L} = \frac{1}{2}\|{\bm y}_{\rm p} - {\bm S}{\bm \gamma}\|_2 + \nu\|{\bm \gamma}\|_1,
\end{align}
where ${\bm S} = [{\bm s}_1,\ldots,{\bm s}_K]$, and $\nu$ is set to $0.05$. 
In the AMP and LASSO frameworks, device activity is determined based on whether the estimated channel gain is larger than a predefined threshold.
The FCNN framework involves two distinct DNNs, each performing CE and AD tasks. These networks comprise $10$ hidden layers with $2048$ neurons per layer and ReLU activation functions. Other training parameters are the same as those of our framework. In the coherent scheme, all frameworks employ the MRC method in \eqref{eq:est_data} to detect the symbols $\{x_{k,d}\}_{\forall k,d}$. 
For comparison, in the non-coherent scheme, we modify the AMP and LASSO frameworks to ensure the constraint in \eqref{eq:true_akj_NC}, similar to \cite{CS1,Model_driven}. We refer to these modified frameworks as Non-coherent AMP and Non-coherent LASSO, respectively.

    
\begin{figure*}[t]
    \centering 
    \subfigure[Average error probability]
    {\epsfig{file=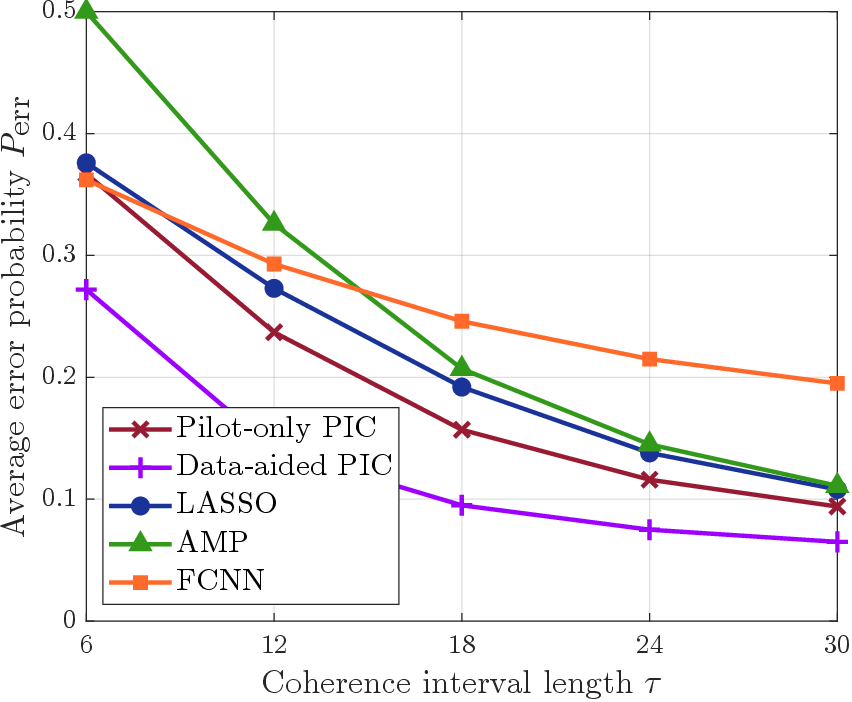, width=6.0cm}} 
    \subfigure[NMSE]
    {\epsfig{file=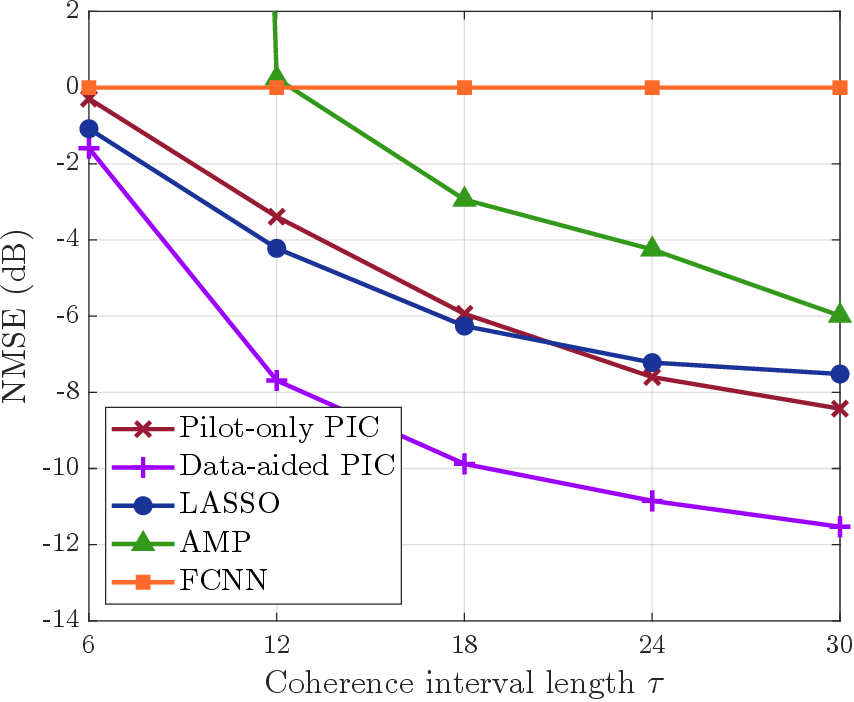, width=6.0cm}} 
    \subfigure[BER]
    {\epsfig{file=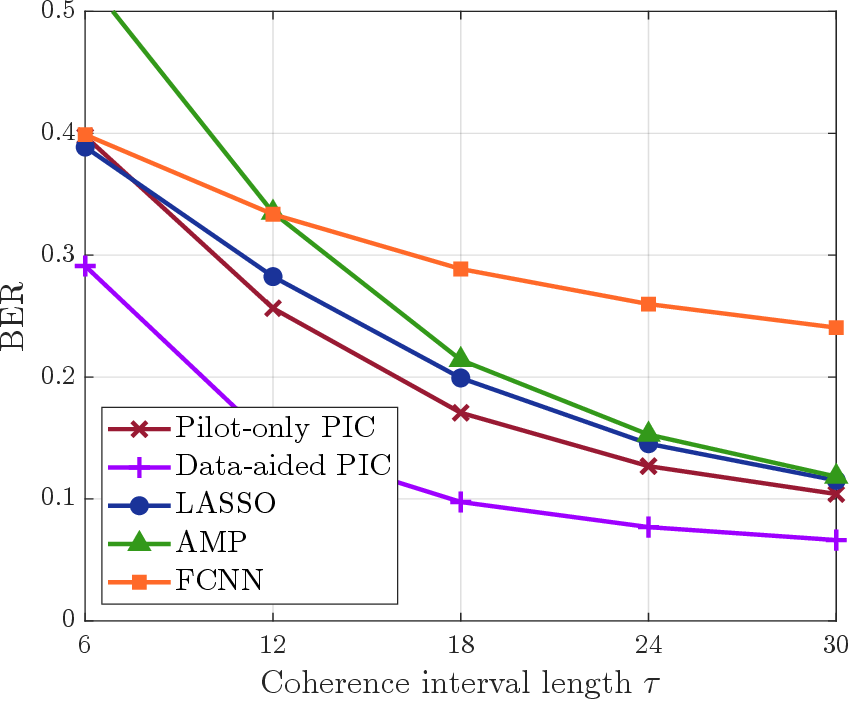, width=6.0cm}} 
    \caption{Performance comparison of different grant-free NOMA frameworks with the coherent scheme for various values of $\tau$.} 
    \label{fig:C}
\end{figure*}

\subsection{Evaluation with Coherent Scheme}
In Fig.~\ref{fig:P_MD_P_FA}, we compare the false alarm probability $P_{\rm fa}$ and miss detection probability $P_{\rm md}$ of different grant-free NOMA frameworks with the coherent scheme for various values of $\epsilon$.  
For the proposed frameworks, we set $L=6$, $T=4$ for the pilot-only PIC, and $T=2$ for the data-aided PIC. 
Their simulation times are comparable but significantly lower than those of the other frameworks. 
Fig.~\ref{fig:P_MD_P_FA} shows that the proposed pilot-only and data-aided PIC frameworks outperform all existing frameworks regardless of $\epsilon$, even though they were trained with a fixed $\epsilon=0.25$. 
In particular, the performance gap between the proposed and existing frameworks increases as $\epsilon$ decreases. 
This result not only highlights the superiority of the proposed frameworks but also demonstrates their effective applicability to diverse grant-free NOMA scenarios. 
Fig.~\ref{fig:P_MD_P_FA} also shows that the proposed data-aided PIC framework achieves the lowest error rates for all $\epsilon$. This result suggests that the data-aided PIC framework effectively extracts the common channel information shared between the received pilot and data signals. 
Meanwhile, our frameworks exhibit higher AD accuracy compared to the LASSO framework. This improvement can be attributed to our optimization strategy which relies on the joint loss function in \eqref{eq:loss_all}, differing from the loss function in \eqref{eq:loss_LASSO}. 



In Fig.~\ref{fig:C}, we compare the average error probability $P_{\rm err}$, NMSE, and BER of different grant-free NOMA frameworks with the coherent scheme for various values of $\tau$. In this simulation, we set $\epsilon=0.1$, $J=2$, $L=\tau/3$, and determine appropriate thresholds to ensure the condition $P_{\rm err} = P_{\rm fa} = P_{\rm md}$. The BER is then evaluated under this specified condition. 
Fig.~\ref{fig:C} shows that $P_{\rm err}$, NMSE, and BER of all frameworks decrease as $\tau$ increases. 
Fig.~\ref{fig:C} also demonstrates that the proposed pilot-only and data-aided PIC frameworks outperform other frameworks across all performance metrics. 
The performance gaps between the AMP and proposed frameworks significantly increase as $\tau$ decreases. 
Under this result, we note that the CS-based AMP technique should satisfy the RIP condition to ensure the recovery performance. 
Meanwhile, the performance gaps between the FCNN and the proposed frameworks indicate that incorporating communication knowledge from the PIC into DL is more effective than treating DL as a black box. In particular, the NMSE results demonstrate that it is difficult to optimize a naive FCNN without knowledge of the communication.  

\begin{figure}[t]
    \centering 
    \subfigure[Average error probability]
    {\epsfig{file=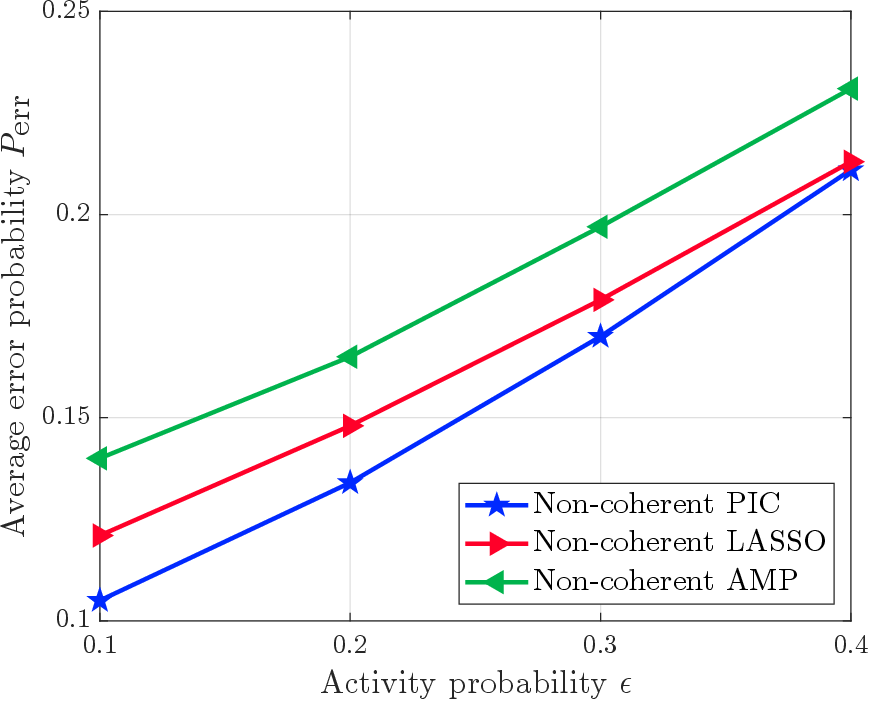, width=6.5cm}} 
    \subfigure[BER]
    {\epsfig{file=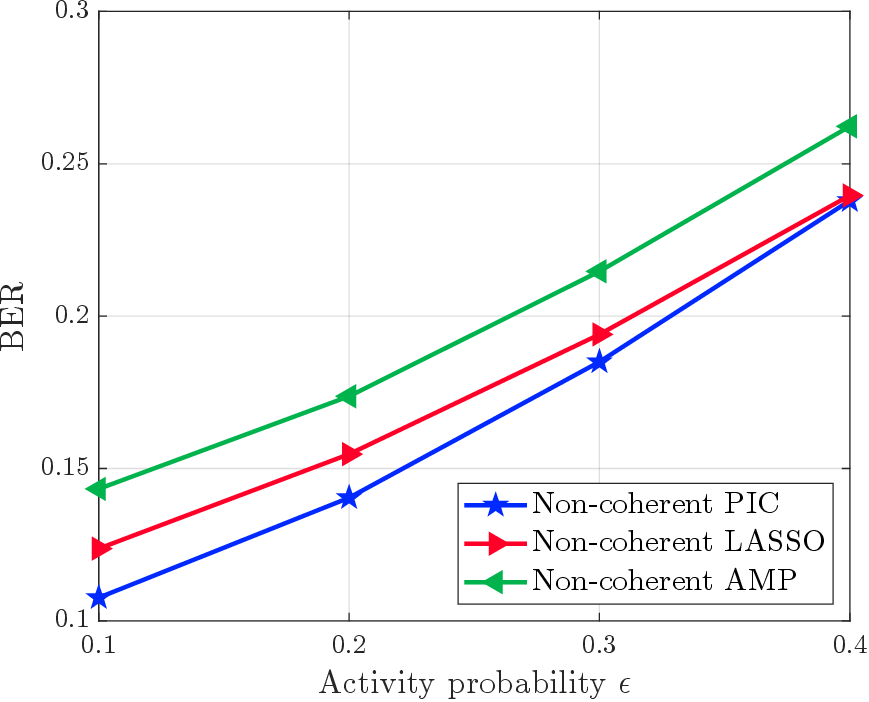, width=6.5cm}} 
    \caption{Performance comparison of different grant-free NOMA frameworks with the non-coherent scheme for various values of $\epsilon$.} 
    \label{fig:NC}
\end{figure}

\begin{figure}[t]
    \centering 
    \subfigure[Average error probability]
    {\epsfig{file=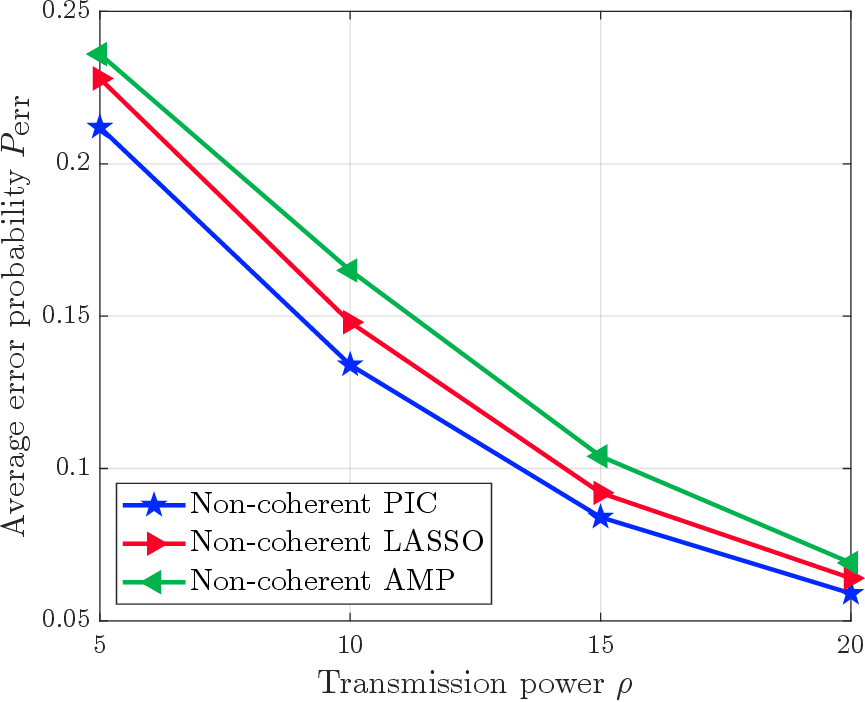, width=6.5cm}} 
    \subfigure[BER]
    {\epsfig{file=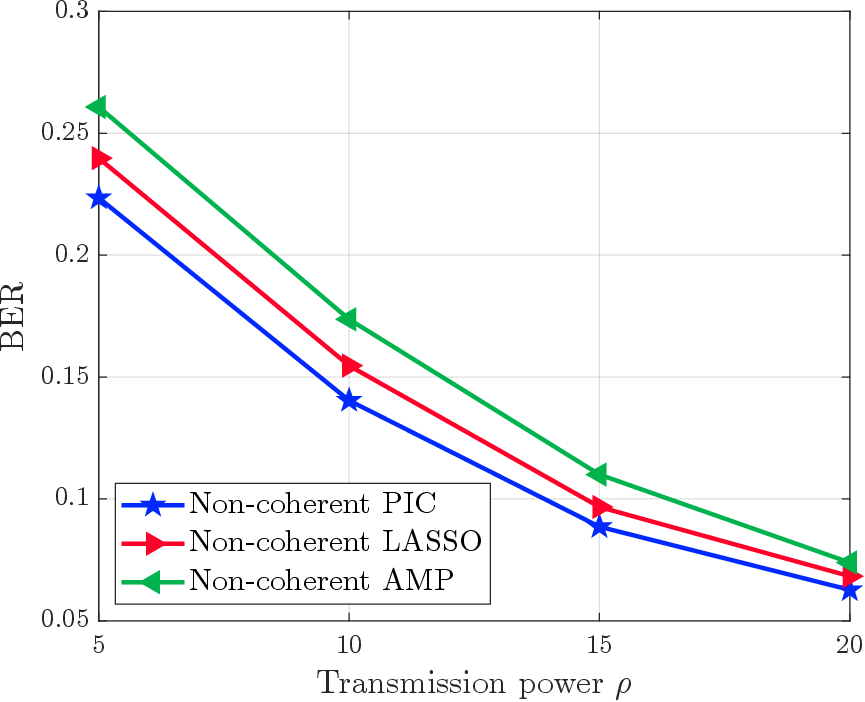, width=6.5cm}} 
    \caption{Performance comparison of different grant-free NOMA frameworks with the non-coherent scheme for various values of $\rho$.} 
    \label{fig:NC_rho}
\end{figure}

\subsection{Evaluation with Non-Coherent Scheme}
In Fig.~\ref{fig:NC}, we compare $P_{\rm err}$ and BER of different grant-free NOMA frameworks with the non-coherent scheme for various values of $\epsilon$. In this simulation, we set $\tau = 10$, $K=10$, $T=4$, $J=2$, and $\rho=10$ dBm. Fig.~\ref{fig:NC} shows that the proposed non-coherent PIC framework outperforms the existing frameworks. In particular, the performance gap in terms of both $P_{\rm err}$ and BER between the proposed and existing frameworks increases as $\epsilon$ decreases, even though it was trained with a fixed $\epsilon=0.25$. 

In Fig.~\ref{fig:NC_rho}, we compare $P_{\rm err}$ and BER of different grant-free NOMA frameworks with the non-coherent scheme for various values of transmission power $\rho$. In this simulation, we set $\epsilon=0.2$, $\tau = 10$, $K=10$, $T=4$, and $J=2$. 
Fig.~\ref{fig:NC_rho} shows that the proposed framework achieves the lowest $P_{\rm err}$ and BER across varying $\rho$. This result suggests the robustness of the proposed framework for various values of $\rho$.  


\begin{table}[t]
    \renewcommand{\arraystretch}{1.5}
    \caption{BER and $P_{\rm err}$ of different grant-free NOMA frameworks for various values of $J$.}\label{table:C_vs_NC}
    \small
    \centering
    \begin{tabular}{|c|c|c|c|c|} \hline
        & \multicolumn{2}{c|}{BER} & \multicolumn{2}{c|}{$P_{\rm err}$} \\ \hline 
        The number of bits $J$ & $1$ & $2$ & $1$ & $2$ \\ \hline\hline
        Pilot-only PIC & ${\bf 0.123}$ & ${\bf 0.263}$ & ${\bf 0.101}$ & ${\bf 0.230}$ \\\hline
        Data-aided PIC & ${\bf 0.063}$ & ${\bf 0.120}$ & ${\bf 0.062}$ & ${\bf 0.114}$ \\ \hline
        LASSO & $0.167$ & $0.294$ & $0.150$ & $0.274$ \\ \hline
        AMP & $0.178$ & $0.324$ & $0.162$ & $0.304$ \\ \hline
        FCNN & $0.242$ & $0.327$ & $0.147$ & $0.242$ \\ \hline\hline
        Non-coherent PIC & ${\bf 0.055}$ & ${\bf 0.089}$ & ${\bf 0.053}$ & ${\bf 0.084}$ \\ \hline
        Non-coherent LASSO & $0.067$ & $0.096$ & $0.064$ & $0.092$ \\ \hline
        Non-coherent AMP & $0.068$ & $0.110$ & $0.066$ & $0.104$\\ \hline
    \end{tabular}
\end{table}

\subsection{Comparison between Coherent and Non-Coherent Schemes}
In Table~\ref{table:C_vs_NC}, we compare the coherent and non-coherent schemes using different grant-free NOMA frameworks. In this simulation, we set $\epsilon=0.2$, $\tau = 10$, $K=10$, and $\rho=15$ dBm. For the proposed frameworks, we set $T=7$ for the pilot-only PIC, $T=3$ for the data-aided PIC, and $T=4$ for the non-coherent PIC. Their simulation times are comparable but significantly lower than those of the other frameworks. 
It should be noted that the sequence length is determined differently between the coherent and non-coherent schemes under the fixed $\tau=10$. Specifically, for the coherent scheme, it is calculated as $L=\lfloor \tau/(J+1)\rfloor$, and for the non-coherent scheme, it is set to $L=\tau=10$. 
Table~\ref{table:C_vs_NC} shows that the grant-free NOMA frameworks with the non-coherent scheme achieve lower BER compared to those with the coherent scheme. In particular, the proposed non-coherent PIC framework achieves the lowest error rates. 
This result suggests that the non-coherent scheme is particularly useful for scenarios transmitting a small number of data bits \cite{CS1}. 



\begin{table}[t]
    \renewcommand{\arraystretch}{1.5}
    \caption{Performance of different grant-free NOMA frameworks with $50000$ test data samples when $\epsilon=0.1$ and $\tau=10$.}\label{table:Sim_time}
    \small
    \centering
    \begin{tabular}{|c|c|c|c|} \hline
        & $P_{\rm err}$ & BER & Simulation time (s) \\ \hline\hline
        Pilot-only PIC-I & \multirow{3}{*}{$0.179$} & \multirow{3}{*}{$0.194$} & $0.018$ \\ \cline{1-1} \cline{4-4}
        Pilot-only PIC-II &  & & $0.750$ \\ \cline{1-1} \cline{4-4}
        Pilot-only PIC-III &  & & $264.176$ \\ \hline
        Data-aided PIC-I & \multirow{3}{*}{$0.091$} & \multirow{3}{*}{$0.093$} & $0.016$ \\ \cline{1-1} \cline{4-4}
        Data-aided PIC-II &  & & $0.703$ \\ \cline{1-1} \cline{4-4}
        Data-aided PIC-III &  & & $258.518$ \\ \hline
        FCNN-I & \multirow{3}{*}{$0.203$} & \multirow{3}{*}{$0.264$} & $0.006$ \\ \cline{1-1} \cline{4-4}
        FCNN-II &  &  & $12.988$ \\ \cline{1-1} \cline{4-4}
        FCNN-III &  &  & $389.118$ \\ \hline
        LASSO & $0.217$ & $0.224$ & $848.751$ \\ \hline
        AMP & $0.248$ & $0.255$ & $394.774$ \\ \hline\hline
        Non-coherent PIC-I & \multirow{3}{*}{$0.106$} & \multirow{3}{*}{$0.108$} & $0.021$ \\ \cline{1-1} \cline{4-4}
        Non-coherent PIC-II &  & & $1.103$ \\ \cline{1-1} \cline{4-4}
        Non-coherent PIC-III &  & & $272.600$ \\ \hline
        Non-coherent LASSO & $0.121$ & $0.124$ & $981.684$ \\ \hline
        Non-coherent AMP & $0.140$ & $0.143$ & $397.015$ \\ \hline
    \end{tabular}\vspace{-2mm}
\end{table}

In Table~\ref{table:Sim_time}, we evaluate $P_{\rm err}$, BER, and simulation time\footnote{Simulation time is measured using Intel(R) Core(TM) i7-10700K CPU @ 3.80 GHz, NVIDIA GeForce RTX 3070, and 32.0 GB RAM.} of different grant-free NOMA frameworks, using $50,000$ test data samples. In this simulation, we set $\epsilon = 0.1$, $\tau=10$, $K=10$, and $\rho=10$ dBm. 
For both the FCNN and proposed frameworks, we consider three data sample processing scenarios: (i) employing GPU to process all data samples at once, (ii) utilizing CPU to process all data samples at once, and (iii) using CPU to process each data sample. 
For each scenario, we use the notations FCNN-$j$ and Proposed-$j$, where $j\in\{{\rm I}, {\rm II}, {\rm III}\}$ for notational simplicity. 
For conventional frameworks, we consider the third scenario due to the iterative nature of their algorithms. 
Table~\ref{table:Sim_time} demonstrates that the proposed frameworks achieve the lowest error rates while maintaining short simulation times. 
Although FCNN-I requires less simulation time than PIC-I, the proposed frameworks outperform the FCNN framework in both $P_{\rm err}$ and BER performances by considerable margins. 
Meanwhile, simulation times for the first and second scenarios are significantly lower than for the third scenario in the proposed frameworks. This result implies that the proposed frameworks can be utilized effectively when the BS needs to perform AD and DD from multiple received signals stored in the buffer, a common scenario in real-world applications. 



\section{Conclusion}\label{Sec:Conclusion}
In this paper, we proposed a novel DL-assisted PIC approach for joint AD, CE, and DD tasks for the uplink grant-free NOMA system. 
We have tackled the inherent challenges in existing grant-free NOMA frameworks, including performance degradation due to a limited spreading sequence length, non-orthogonal spreading sequences, and lack of knowledge of channel gain orders and device activities, by leveraging the advantages of PIC and DL. 
Specifically, we have developed three PIC-based frameworks, namely, the pilot-only PIC, the data-aided PIC, and the non-coherent PIC frameworks.  
To enhance training efficiency and mitigate the vanishing gradient problem, we designed the end-to-end training strategy facilitated by joint loss function and IC modules. This strategy not only enables joint optimization for AD, CE, and DD tasks but also allows for training complex relationships between different modules, leading to improved performance. An important direction of future research is extending our framework to incorporate spreading sequence design, aiming to preserve orthogonality and alleviate interference. 

\bibliographystyle{IEEEtran}
\bibliography{Reference}

\end{document}